\documentclass[a4paper,11pt,showpacs]{revtex4}
\usepackage{amssymb,amsmath}

\usepackage[british]{babel}
\usepackage{graphicx}

\newcommand{\diff}{\mathrm{d}}
\begin{document}

\title{SLE on doubly-connected domains and the winding of loop-erased random walks}

\author{Christian Hagendorf}
\email{hagendor@lpt.ens.fr}
\author{Pierre Le Doussal}
\email{ledou@lpt.ens.fr}
\affiliation{CNRS-Laboratoire de Physique Th\'eorique de
l'Ecole Normale Sup\'erieure\\24, rue Lhomond, 75231
Paris cedex 05, FRANCE\\
}

\begin{abstract}
Two-dimensional loop-erased random walks (LERWs) are random planar curves whose scaling limit is known to be a Schramm-Loewner evolution SLE$_{\kappa}$ with parameter $\kappa=2$. In this note, some properties of an SLE$_\kappa$ trace on doubly-connected domains are studied and a connection
to passive scalar diffusion in a Burgers flow is emphasised. In particular, the endpoint probability distribution and winding probabilities for SLE$_2$ on a cylinder, starting from one boundary component and stopped when hitting the other, are found. A relation of the result to conditioned one-dimensional Brownian motion is pointed out. Moreover, this result permits to study the statistics of the winding number for SLE$_2$ with fixed endpoints. A solution for the endpoint distribution of SLE$_4$ on the cylinder is obtained and a relation to reflected Brownian motion pointed out.
\end{abstract}

\date{\today}

\maketitle

\section{Introduction}

The Schramm-Loewner Evolution SLE$_\kappa$ is a one-parameter family of conformally invariant measures on non-intersecting planar curves. In many cases, interfaces in two-dimensional models of statistical mechanics at criticality are conjectured to be described by SLE. For example, the scaling limit of the planar loop-erased random walk (LERW) is known to be SLE$_2$, interfaces in the 2D Ising model SLE$_3$ and critical percolation hulls to be described by SLE$_6$. While these systems can be also studied using
traditional methods of theoretical physics like (boundary) conformal field theory and Coulomb gas methods, Schramm has given a novel and rigourous approach via conformally invariant stochastic growth processes.

LERWs were the starting point and motivation for the development of SLE in Schramm's seminal paper \cite{schramm:00}. Today their scaling limits and relations to SLE are well understood in simply-connected domains, where conformal invariance was shown by Lawler, Schramm and Werner \cite{lsw:04}.
Properties in multiply-connected domains are more involved but much recent progress was obtained, e.g. in a series of works by Dapeng Zhan \cite{zhan:04, zhan:06, zhan:06_2}.
In this paper, we mainly study properties of planar LERWs/SLE$_2$ in doubly-connected domains, emerging from one boundary component and conditioned to finish when hitting the second boundary component. In contrast to the simply-connected case, curves in doubly-connected domains may wind non-trivially around various boundary components. Here we concentrate on these winding properties of loop-erased random walks on cylinders in two cases: (1) with one endpoint fixed on one boundary component, and the other endpoint free on the other boundary component, (2) with both endpoints fixed on different boundary components. We establish these results by solving a differential equation arising from SLE$_\kappa$ with $\kappa=2$ on the cylinder. Another motivation to study the cylinder is that numerical simulations of critical systems are often performed in a cylinder geometry, see e.g. \cite{alan}. 

The paper is organised as follows. In section $2$ we give a brief description of SLE. In particular, we outline two variants of SLE in simply-connected domains which prove to be limits (in some sense) of the doubly-connected case. After this we discuss SLE on cylinders, being prototypes of doubly-connected domains. Section $3$ provides a detailed discussion on endpoint probabilities and winding properties of SLE$_\kappa$ on cylinders. Inspired from analogies with diffusion-advection in $(1+1)$-dimensional Burgers flows, we present analytical results for the case $\kappa=2$, study various limits and point out a relation between the winding of LERWs and conditioned one-dimensional Brownian motion. Finally, conditioning walks to exit via a given point, we derive the winding statistics for LERW with fixed endpoints on the cylinder. In section $4$, somewhat aside the main topic, we provide the endpoint distribution in the case $\kappa=4$ and point out a relation to reflected two-dimensional Brownian motion. We conclude in section $5$. The appendix contains computations of exit probabilities for Brownian motion and random walks on the lattice,
in relation to distribution of exit points for SLE$_2$ and their finite-size corrections. It also contains a partial (periodised) result about bulk left-passage probabilities with respect to a point in the covering space of the cylinder, as well as a path-integral version of the endpoint distribution problem. 

\section{Basic notions of SLE}
In this section we give an informal description of SLE. For detailed accounts we refer the reader to the many existing excellent review articles on this subject (see e.g. \cite{cardy:05,bauer:06,lawler:06}).

Stochastic Loewner evolutions (SLE) describe the growth of
non-inter{\-}secting random planar curves or hulls emerging from the
boundary of planar domains $\mathbb{D}$ by means of stochastic
differential equations. A crucial feature of SLE is \textit{conformal invariance}: if two domains $\mathbb{D}$ and $\mathbb{D}'$ are related by a conformal mapping $g(z)$, then the image of SLE in $\mathbb{D}$ is SLE in $\mathbb{D'}$. Another key property is the so-called \textit{domain Markov property} of SLE. Conditioning on the curve/hull $\gamma_t$ grown up to time $t$ in $\mathbb{D}$, the remainder of the curve is just SLE in the slit domain $\mathbb{D}_t=\mathbb{D}\backslash \gamma_t$. In fact, these two properties are (almost) imply that the growing hulls a characterised by a single positive parameter $\kappa$.

For simply-connected domains, conformal invariance and the Riemann mapping theorem naturally lead to a description of the curves in terms of conformal mappings $g_t(z)$ which uniformise $\mathbb{D}_t$ to $\mathbb{D}$, and map the tip of $\gamma_t$ to some point on the boundary of $\mathbb{D}$. $g_t(z)$ is solution of a differential equation in $t$ with initial condition $g_{t=0}(z)=z$, called the \textit{Loewner equation}.
SLE in simply-connected domains comes in different flavours, according to which the Loewner differential equation varies. In section \ref{sec:slescd}, we shall concentrate on so-called radial and dipolar SLEs.

Compared to simply-connected domains, SLE in multiply-connected
domains, proves to be more involved because no equivalent of the Riemann mapping theorem is at hand \cite{zhan:04, zhan:06,bauer:04, bauer:04_2}. In this article, we concentrate on doubly-connected domains $\mathbb{D}$ with the topology of an annulus. Their boundary is given by two disjoint simple curves $\partial_1\mathbb{D}$ and $\partial_2\mathbb{D}$. A classical theorem of complex analysis states that each such a domain may be mapped onto the annulus $\mathbb{A}_p=\{z\in \mathbb{C}\mathop{|} e^{-p}<|z|<1\}$ with some $p>0$, such that for example $\partial_1\mathbb{D}$ is mapped to $|z|=1$ and $\partial_2\mathbb{D}$ to $|z|=e^{-p}$. Conformal invariance of SLE legitimates the choice of $\mathbb{A}_p$ as a reference domain. However, we shall rather work with a cylinder $\mathbb{T}_p = \{z=x+iy\in \mathbb{C}\mathop{|} -\pi < x \leq \pi,\, 0<y<p\}$, identifying $x=-\pi$ and $x=\pi$.  The cylinder $\mathbb{T}_p$ may always be mapped back onto $\mathbb{A}_p$ via the conformal mapping $w = e^{iz}$. The height of the cylinder $p>0$ is called \textit{the modulus}, and the boundary components are given by $\partial_1\mathbb{T}_p = [-\pi,\pi]$ and $\partial_2\mathbb{T}_p = [-\pi,\pi]+ip$ which we shall refer to as ``lower'' and ``upper boundary'' respectively.  We shall study hulls $\gamma_t$ growing from $z = 0 \in \partial_1\mathbb{T}_p$ to $\partial_2\mathbb{T}_p$ as shown on figure \ref{fig:strip}a and b. In section \ref{sec:sledcd}, we analyse the Loewner equation for $g_t(z)$ mapping $\mathbb{T}_p\backslash \gamma_t$ to some cylinder $\mathbb{T}_q$ with (in general) different modulus $q\neq p$.  A mapping back to $\mathbb{T}_p$ does not exist since one may show that there is no conformal mapping sending a cylinder of modulus $p$ to a cylinder of modulus $q\neq p$ \cite{nehari:82}. Moreover, we show how the doubly-connected version of SLE on $\mathbb{T}_p$ interpolates between radial and dipolar SLE.

\subsection{SLE in simply-connected domains}
\label{sec:slescd}

Before embarking into considerations on general values for the modulus $p$ we shall consider limiting cases of semi-infinite cylinders $p\to +\infty$ and very thin cylinders $p\to 0^+$ in order to establish a connection with well-known variants of SLE in simply-connected domains.

\medskip

\textit{$p\to +\infty$ and radial SLE.} Radial SLE starts from a boundary point $x_0\in \partial \mathbb{D}$ to a bulk point $x_\infty\in \mathbb{D}$ of some simply-connected planar domain $\mathbb{D}$. Its Loewner equation is most conveniently written for the unit disc $\mathbb{D}=\mathbb{U}$ from $x_0=1$ to $x_\infty=0$ where
\begin{equation} \label{radialeq}
  \frac{\diff g_t(z)}{\diff t} =-g_t(z)\frac{g_t(z) + e^{i\xi_t}}{g_t(z) - e^{i\xi_t}},
\end{equation}
where $\xi_t$ is some real-valued function of $t\geq 0$ and $g_t(0)=0$ (the point $x_\infty$ remains unchanged during time evolution).
The time parametrisation has been chosen so that $g'_t(0)=e^t$. The real-valued function $\xi_t$ is related to the image of the tip $\tau_t$ of the growing trace $\gamma_t$ by $g_t(\tau_t) = e^{i\xi_t}$. In fact, conformal invariance and the domain Markov property imply that $\xi_t = \sqrt{\kappa}B_t+\alpha t$ is simple Brownian motion with a diffusion constant $\kappa$ and a linear drift $\alpha t$. If we ask for rotation invariance as it occurs in physical systems (and throughout this paper), the drift vanishes $\alpha=0$.  For later considerations, we outline another useful version of radial SLE. In fact, the unit disc $\mathbb{U}$ may be sent to the semi-infinite cylinder $\mathbb{T}_\infty$ via the conformal mapping $w=-i \log z$ (here the principal branch of the logarithm has been chosen). In this geometry, the Loewner equation for the new uniformising map $\tilde{g}_t(z)$ is given by
\begin{equation}
  \frac{\diff \tilde g_t(z)}{\diff t} = \cot\left(\frac{\tilde g_t(z)-\xi_t}{2}\right), \qquad\xi_t = \sqrt{\kappa}B_t.
  \label{eqn:radialcyl}
\end{equation}
obtained from (\ref{radialeq}) using $g_t(z)=\exp[{i \tilde g_t(w(z))}]$.

Since we aim at studying winding probabilities, let us recall how to estimate the distribution of the winding angle $\theta_t$ for large $t$ \cite{schramm:00,cardy:05}.  For the unit disc $\mathbb{U}$ as $t\gg 1$, the tip of the radial SLE trace $\tau_t$ approaches $z=0$. We may therefore write $g_t(\tau_t)=e^{i\xi_t}\approx e^t\tau_t$, by using the Taylor series expansion $g_t(z)= e^t z+\dots$ near $z=0$, from what follows $\tau_t \approx e^{-t+i\xi_t}$. Hence the winding angle $\theta_t = \arg \tau_t = \xi_t$ turns out to be a Gaussian random variable with zero mean and variance $\kappa t$. For the infinite cylinder $\mathbb{T}_\infty$, the winding angle corresponds to the real part of the tip allowed to vary continuously on $\mathbb{R}$, and its probability distribution is approximately given by
\begin{equation}
  \lambda_r(t,x) \approx \frac{1}{\sqrt{2\pi\kappa t}}\exp\left(-\frac{x^2}{2\kappa t}\right), \quad \text{for large }t.
  \label{eqn:radialepd}
\end{equation}

\medskip

\textit{$p\to 0^+$ and dipolar SLE.} Another version of SLE is obtained by considering planar curves in a simply-connected domain $\mathbb{D}$ from a boundary point $x_0\in \partial \mathbb{D}$ to a side arc $S\subset \partial \mathbb{D}$ \cite{bauer:05} not containing $x_0$. Let us choose the geometry of an infinite strip $\mathbb{S}_p = \{ z\in \mathbb{C}\mathop{|} 0<\text{Im}\,z<p\}$ of height $p$, $x_0=0$ and $S=\mathbb{R}+ip$. Then Loewner's equation is given by
\begin{equation}
  \frac{\diff g_t(z)}{\diff t} =\frac{\pi}{p}\coth\left[\frac{\pi(g_t(z) - \xi_t)}{2p}\right].
\end{equation}
$g_t(z)$ maps the tip of the trace $\gamma_t$ to $\xi_t$. Again, because of conformal invariance and the domain Markov property we have $\xi_t = \sqrt{\kappa}B_t$.
The trace $\gamma_t$ hits the boundary arc $\mathbb{R}+ip$ at some (random) point $x+ip$ as $t\to +\infty$. 
This version of SLE is related to the cylinder problem on $\mathbb{T}_p$ in the limit $p\to 0^+$. Loosely spoken, if $p$ is very small the SLE trace will not feel the periodicity of the cylinder but only hit the upper boundary close to the point $ip$ on the upper boundary (this statement can be made more precise, see below). The equivalent to the winding angle distribution for radial SLE is the probability distribution of this exit point for dipolar SLE, which we shall denote $\lambda_d(p,x)$. For any $\kappa$ it is given by \cite{bauer:05}
\begin{equation}
  \lambda_d(p,x) = \frac{\sqrt{\pi}\,\Gamma(1/2+2/\kappa)}{2p\,\Gamma(2/\kappa)}\frac{1}{\displaystyle[\cosh\left(\pi x/2p\right)]^{4/\kappa}}
  \label{eqn:dipolarepd}
\end{equation}

\subsection{Stochastic Loewner evolution on doubly-connected domains}
\label{sec:sledcd}
Let us now turn to the case of general modulus $p$. The Loewner equation for the uniformising map $g_t(z)$ of
SLE$_{\kappa}$ on $\mathbb{T}_p$ is given by \cite{zhan:06}
\begin{equation}
  \frac{\diff g_t(z)}{\diff t} = H\big(p-t,g_t(z)-\sqrt{\kappa}\,B_t\big)
  \label{eqn:loewner}
\end{equation}
with standard Brownian motion $B_t$ and the function
\begin{equation}
  H(p,z) = \lim_{n\to\infty}\sum_{k=-n}^{n}\cot\left(\frac{z}{2}-ikp\right)=\cot \frac{z}{2}+\sum_{k=1}^{\infty}\frac{2\sin z}{\cosh 2kp - \cos z}  \label{eqn:vf}
\end{equation}
The process is defined between $t=0$ ($g_0(z)=z$) where the trace
starts from point $z=0$ on the lower boundary up to $t=p$ where the
SLE trace hits the upper boundary at some random point $\gamma_p$ with $\text{Im}\,\gamma_p = p$. The time parametrisation of the curves is chosen in such a way that  $g_t(z)$ maps $\mathbb{T}_p\backslash \gamma_t$ to $\mathbb{T}_q$ with $q=p-t$. One possible way to derive \eqref{eqn:loewner} and \eqref{eqn:vf} is by considering an infinitesimal hull, and using invariance properties of $\mathbb{T}_p$, namely translation invariance $z\mapsto z+a,\,a\in\mathbb{R}$ and reflection invariance $z\mapsto ip-z$ (see \cite{bauer:04_2} where this has been explicitly carried out for an annulus geometry).
\begin{figure}[htbp]
\begin{center}
\begin{tabular}{cc}
\includegraphics[width=0.35\textwidth]{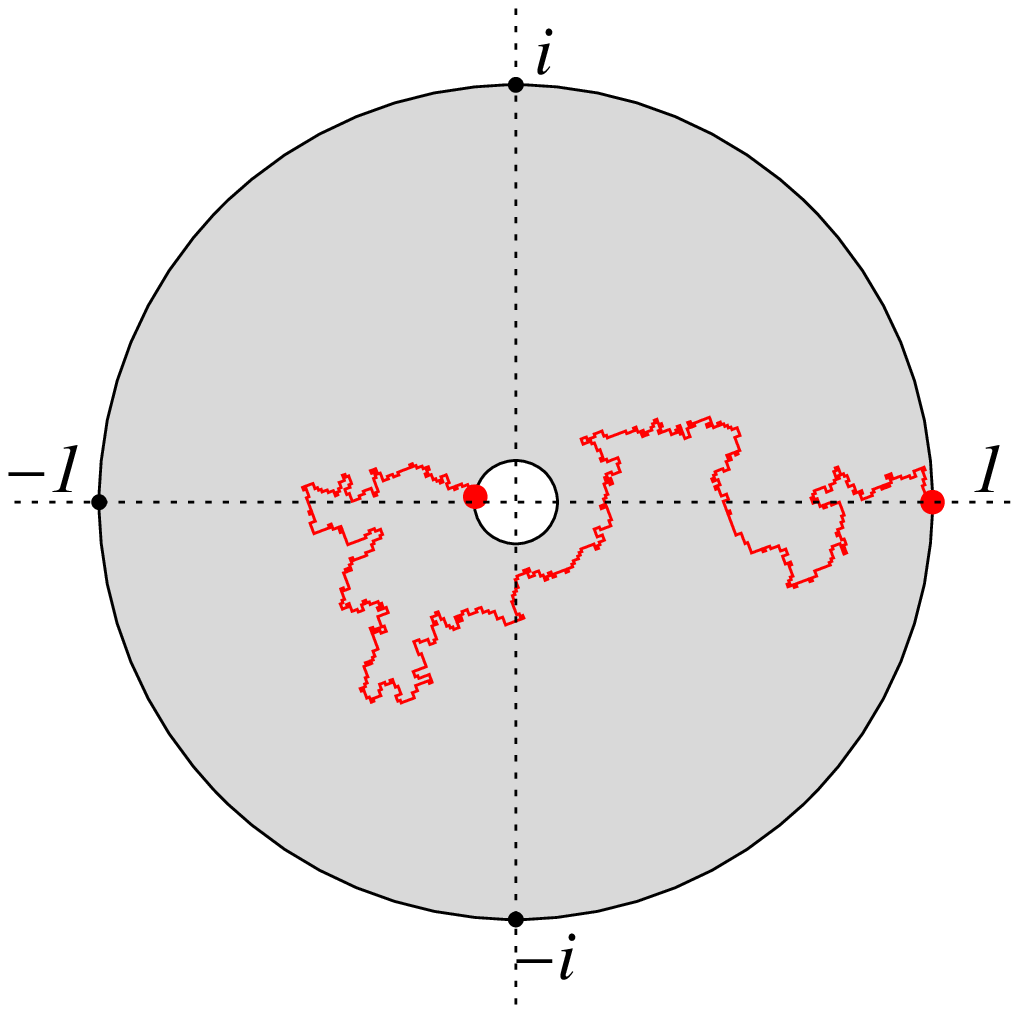}&
\includegraphics[width=0.45\textwidth]{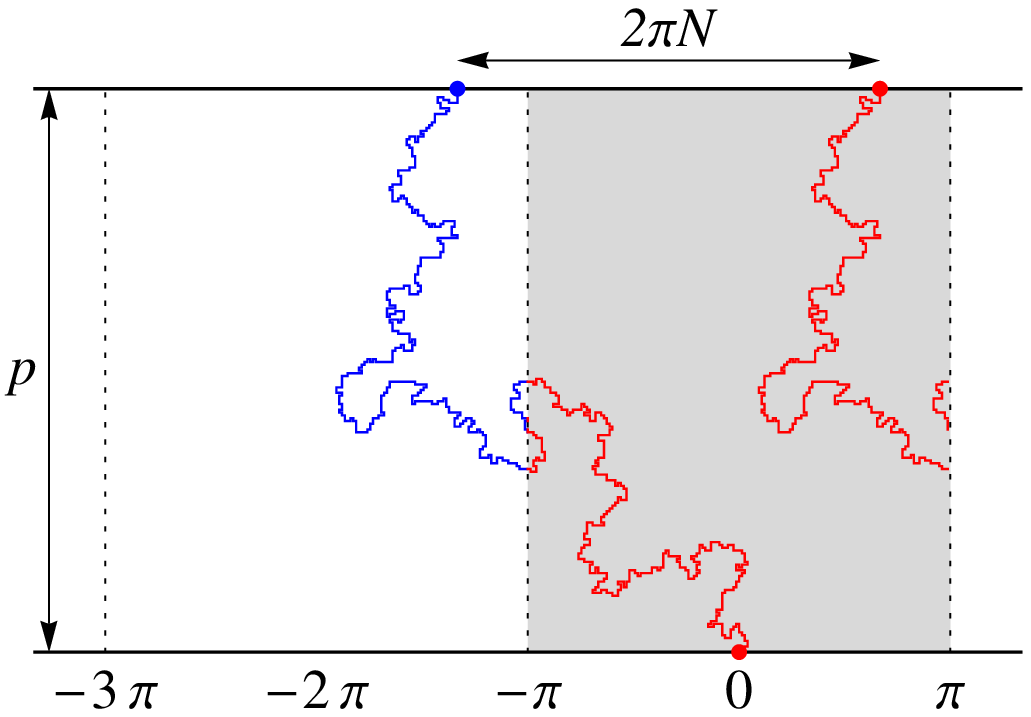}\\
(a) & (b)
\end{tabular}
\end{center}
\caption{(a) Illustration for SLE on an annulus $\mathbb{A}_p$ from $1$ to the target circle $|z|=e^{-p}$. (b) SLE on the cylinder and the covering space: the shaded region represents the cylinder, with sides $x=-\pi$ and $x=\pi$ identified. We obtain the covering space by periodic extension along the real axis. The endpoints on the cylinder and the covering space differ by $2\pi N$ where $N$ denotes the winding number of the trace. The interfaces are samples of loop-erased random walks.}
\label{fig:strip}
\end{figure}

Let us describe the properties of the function defined in \eqref{eqn:vf}.
$H(p,z)$ is an odd function in $z$ with period $2\pi$. Furthermore
it is quasi-periodic in the imaginary direction: $H(p,z+2 i p  ) =
H(p,z)-2i$. Moreover, notice the interesting property $H(p,z)=(i\pi H(\pi^2/p,i\pi z/p)-z)/p$. It may be used in order to derive an alternative series expansion
\begin{equation}
  H(p,z)=\frac{\pi}{p}\lim_{n\to\infty}\sum_{k=-n}^n\coth\left(\frac{\pi(z+2\pi k)}{2p}\right)-\frac{z}{p}.
\end{equation}
For large positive $p$, we have $H(p-t,z) = \cot (z/2)+\dots$ as it can be seen from \eqref{eqn:vf}. Therefore, as we let $p\to+\infty$, the Loewner equation \eqref{eqn:loewner} reduces to the one for radial SLE, defined in \eqref{eqn:radialcyl}.  The dipolar limit $p\to 0^+$ is more involved. Consider the function $h_t(z)=g_t(z)-\sqrt{\kappa}B_t$ mapping $\mathbb{T}_p\backslash \gamma_t$ to $\mathbb{T}_{p-t}$ and the tip $\tau_t$ of the curve to $0$. Rescaling $k_s(z) = \pi\, h_t(z)/(p-t)$ maps the problem to a cylinder of fixed height $\pi$, and width $2\pi^2/(p-t)$. The new mapping $k_s(z)$ depends on a time $s=\pi^2/(p-t),\,0\leq t < p$.
Performing the time change leads to a stochastic differential equation
$
  \diff k_s(z) = i H(s,i\pi k_s(z))\diff s-\sqrt{\kappa}\,\diff \tilde{B}_s
$
where $\tilde{B}_s$ is another standard Brownian motion, obtained from the time-change formula.  Now, as $p\to 0^+$ we have $s\to +\infty$ so that we may use $H(s,z)\sim \cot z/2+\dots$. Finally for $K_s(z) = k_s(z)+\sqrt{\kappa}\,\tilde{B}_s$, we find the dipolar Loewner equation
\begin{equation}
  \frac{\diff K_s(z)}{\diff s} = \coth\left(\frac{K_s(z)-\sqrt{\kappa}\,\tilde B_s}{2}\right)
\end{equation}
in the limit $s\to +\infty$.

\medskip

\textit{The covering space.} In the sequel it shall be convenient to lift the processes to the covering space of the cylinder $\mathbb{T}_p$. In fact, this is simply obtained by allowing $\text{Re}\,\gamma_t$ to vary continuously on $\mathbb{R}$. An illustration is given on figure \ref{fig:strip}b. It is easy to see that the difference between the endpoints of a given trace on the cylinder and its analogue on the covering space is always given by $2\pi N$ with an integer $N$. We shall call $N$ the winding number of the SLE trace. For the covering space, the Loewner equation \eqref{eqn:loewner} still holds. However, probabilities and probability densities computed on the covering space have different boundary conditions when compared to the periodic case of cylinders \cite{zhan:06}.

\section{Endpoint probabilities for SLE on the cylinder}

In this section, we characterise the endpoint distributions for SLE on cylinders of height $p$ which we denote $\Lambda(p,x),\, -\pi \leq x \leq \pi$. To be precise, $\Lambda(p,x)\diff x$ corresponds to the probability that the trace hits the upper boundary within $\text{Re}\,\gamma_p\in [x,x+\diff x]$ for some infinitesimal $\diff x$. In order to study the winding of the traces around the cylinder, we introduce an analogous object $\lambda(p,x),\, x\in \mathbb{R}$ for the covering space. Results on the cylinder are obtained upon identification of points differing by $2\pi n$:
\begin{equation}
  \Lambda(p,x) = \sum_{n=-\infty}^\infty \lambda(p,x+2\pi n)
  \label{eqn:periodisation}
\end{equation}

\subsection{Endpoint distributions for general $\kappa$}

In order to compute $\Lambda(p,x)$ consider the SLE trace
$\gamma_{\diff t}$ evolving for an infinitesimal time $\diff t$ on the
cylinder $\mathbb{T}_p$. The uniformising map $g_{dt}(z)$ maps the
remainder of the curve $\gamma_p\backslash \gamma_{\diff t}$ onto
SLE in a cylinder $\mathbb{T}_{p-\diff t}$ starting at
$\sqrt{\kappa}\,\diff B_0$. Under this mapping $g_{\diff t}(z)$ the point $x+ip$
on the upper boundary of $\mathbb{T}_p$ is sent to some point $x'+i(p-\diff t)$
on the upper boundary of $\mathbb{T}_{p-\diff t}$, with
$x'=x+(H(p,x+ip)+i)\diff t$, see \eqref{eqn:loewner}. Using
the Markov property and conformal invariance of the SLE measure we
obtain, upon averaging over the infinitesimal Brownian increment $\diff B_0$, the
equation
\begin{equation}
  \Lambda(p,x)\diff x =\mathsf{E}_{\diff B_0}[\Lambda(p-\diff t, x'-\sqrt{\kappa}\,\diff B_0)\diff x']
  \label{eqn:averagebm}
\end{equation}
An expansion to first order in $\diff t$, using $\mathsf{E}_{\diff B_0}[\diff B_0]=0$ and $\mathsf{E}_{\diff B_0}[\diff B_0^2]=\diff t$, leads to the
Fokker-Planck equation
\begin{equation} 
  \frac{\partial \Lambda(p,x)}{\partial p} = \frac{\partial}{\partial
  x}\left(v(p,x)\Lambda(p,x)\right)+\frac{\kappa}{2}
  \frac{\partial^2\Lambda(p,x)}{\partial x^2}
  \label{eqn:pde}
\end{equation}
where we have introduced the drift function (compare with \eqref{eqn:vf})
\begin{equation}
v(p,x) = H(p,x+ip)+i
= \frac{\pi}{p}\lim_{N\to\infty}\sum_{n=-N}^{N} \tanh
\frac{\pi(x+2\pi n)}{2 p} - \frac{x}{p} \label{eqn:ch}
\end{equation}
Like $H(p,z)$ the drift function $v(p,z)$ is an odd and periodic function with period $2\pi$. Furthermore, it is quasiperiodic in imaginary direction: $v(p,z+2ip) = v(p,z)-2i$. Notice that $\lambda(p,x)$ is a solution of \eqref{eqn:pde} as well, however with vanishing boundary conditions for $x\to \pm\infty$. Leaving aside the
identification of the points $x$ and $x+2\pi$, we ask for the
probability
\begin{equation}
  \omega(p,x) = \int_{-\infty}^x \diff y\, \lambda(p,y)
\end{equation}
that the SLE traces hits the upper boundary to
the left of a given point $x+ip$ on the covering space. It is solution of the convection-diffusion equation
\begin{equation}
  \frac{\partial \omega(p,x)}{\partial p} = v(p,x)\frac{\partial
  \omega(p,x)}{\partial x}+\frac{\kappa}{2}
  \frac{\partial^2\omega(p,x)}{\partial x^2},
  \label{eqn:proba}
\end{equation}
with boundary conditions $\lim_{p\to 0^+}\omega(p,x) = \Theta(x)$, $\lim_{x\to +\infty} \omega(p,x)=1$ and $\lim_{x\to -\infty} \omega(p,x)=0$. Here $\Theta(x)$ denotes the Heaviside function which is
$1$ for $x> 0$, and $0$ otherwise. It is a remarkable property of the function
$v(p,x)$ to be a solution of the $(1+1)$-dimensional Burgers equation
\begin{equation}
  \frac{\partial v(p,x)}{\partial p} = v(p,x) \frac{\partial v(p,x)}{\partial
  x}+\frac{\partial^2v(p,x)}{\partial x^2},
  \label{eqn:pde4v}
\end{equation}
where $p$ plays the role of time and viscosity is set to unity \cite{zhan:06}. In fact, this equation even holds for complex $x$. Furthermore
is can be shown that $H(p,z)$ obeys the same equation by using the relationship between $H(p,z)$ and $v(p,z)$. Hence we interpret
(\ref{eqn:proba}) as the evolution equation of a passive scalar convected and diffusing in
the Burgers flow $- v(p,x)$, also described by a Langevin equation:
\begin{equation}
\diff x_p = - v(p,x_p)\diff p + \sqrt{\kappa} \diff B_p
\end{equation}
The problem of finding SLE probabilities for arbitrary $\kappa$ can thus be
reformulated as finding the density of a passive scalar convected by a (particular) one-dimensional Burgers flow (along $x$). In its general formulation it is a classic
problem \cite{bauer:99, bec:07, nagar:06, drossel:02, ginanneschi:97, ginanneschi:98} characterised by the dimensionless Prandl number $P$ \cite{gorodtsov:99}, with $P=2/\kappa$ for our problem.

By means of the well-known
Cole-Hopf transformation \cite{hopf:50, cole:51} we rewrite
\begin{equation}
  v(p,x)= 2\,\frac{\partial}{\partial x}\, \ln \epsilon(p,x),
  \label{eqn:colehopf}
\end{equation}
where $\epsilon(p,x)$ satisfies the diffusion equation
$\partial \epsilon(p,x)/\partial p = \partial^2 \epsilon(p,x)/\partial x^2$ with
solution
\begin{equation}
\epsilon(p,x) = \sum_{n = - \infty}^{+\infty} (-1)^n \exp\left(i n x - n^2
p\right)= \sqrt{\frac{\pi}{ p}}  \sum_{n = - \infty}^{+\infty}
\exp\left(-\frac{(x- \pi - 2 \pi n)^2}{4 p} \right)
\label{eqn:seriesepsilon}
\end{equation}
corresponding to an initial condition $\epsilon(0,x)= 2 \pi
\sum_{n=-\infty}^{\infty}\delta(x-\pi -2\pi n)$. In the language of
Burgers equation this corresponds to an initial condition at $p=0^+$ with a
periodic set of shocks at positions $x_n^s=2 \pi n$. As the ``time'' $p$
increases the shocks are broadened by the viscosity and acquire a width proportional to
$p$, leading to the smooth periodic function (\ref{eqn:ch}). Finding the
solution of (\ref{eqn:pde}) for general $\kappa$ is a complicated time-dependent problem, and only results in special cases are obtained 
below (see appendix \ref{apppath} for a path-integral formulation). 

In the limit of large modulus $p\gg 1$ the drift term in (\ref{eqn:pde})
vanishes like $v(p,x)\sim 4 e^{-p}\sin x$. Therefore we recover a simple diffusion equation for the endpoint distribution $\lambda(p,x)$ which, in terms of the scaling variable 
$x/\sqrt{p}$, converges to the Gaussian result for radial SLE \eqref{eqn:radialepd}. However if one does not rescale $x$, there are non-trivial corrections and one expects instead
that the cumulants converge at large $p$ to $\kappa$ dependent constants, 
$\langle x^{2 n} \rangle_c - \kappa p\, \delta_{n,2} \to X_{2n,\kappa}$, as shown
explicitly below for $\kappa=2$. Going from $\mathbb{T}_p$ to an annulus  with radii $R_1 > R_2=R_1 e^{-p}$ via the conformal mapping $w=R_1 e^{i z}$ (see figure \ref{fig:strip}a for an illustration) we obtain the winding angle distribution $\theta=x$ for SLE starting from the outer boundary and stopped when first hitting the inner boundary (or the reverse if we admit reversibility of annulus SLE). We expect its cumulants of the winding angle to have the form
\begin{equation}
\langle \theta^{2n}\rangle_c = \kappa \ln(R_1/R_2) \delta_{n,2} + X_{2n,\kappa}
\end{equation}
in the limit of large $R_1/R_2$. This differs from the standard result, also obtained from Coulomb gas methods \cite{duplantier:88, cardy:05} or via the argument given for radial SLE, namely $\langle\theta^2\rangle = (4/g) \ln(R_1/R_2)$ with $4/g=\kappa$ (see \eqref{eqn:radialepd}), by universal constants.
One can compare to the corresponding cumulative distribution of winding angle for the standard two-dimensional Brownian motion with absorbing conditions on the boundary of the annulus
(it can be obtained from dipolar SLE with $\kappa=2$, the direct derivation being recalled in \ref{sec:abc})
\begin{equation}
\omega_b(\theta)= \frac{1}{1+ e^{-  \pi \theta/\ln(R_1/R_2)}} 
\end{equation}
which yields much larger cumulants $\langle{\theta^{2n}}\rangle_c 
\sim  (\ln(R_1/R_2))^{2n}$ with prefactors given below in the context of dipolar SLE.

As $p \to 0^+$ a simple argument allows to recover the dipolar SLE hitting probability \eqref{eqn:dipolarepd} from the form of Burgers shocks, which are well separated in that limit. 
Let us suppose that we may restrict (\ref{eqn:ch}) to a single shock, $v_s(p,x)=-x/p+\pi/p \tanh(\pi x/2p)$, i.e. take an initial
condition $\epsilon(0,x)=2 \pi ( \delta(x+\pi)+\delta(x-\pi))$. This gives
a potential $u(p,x)= 2 \log \epsilon(p,x) = - x^2/2 p + 2 \log \cosh(\pi x/ 2 p)+\text{const.}$ 
which in the scaling region $x \sim p$ strongly confines the diffusing particle described
by (\ref{eqn:pde}) near the origin. One easily checks that the (adiabatic) Gibbs measure
$\lambda_{\mathrm{\tiny adiab}}(x,p) \sim e^{- (2/\kappa) u(p,x)} = C(p) \epsilon(x,p)^{-4/\kappa}
=\lambda_{d}(x,p)$
satisfies (\ref{eqn:pde}) in the region $x \sim p$ (the term $\partial \lambda(p,x)/\partial p$ on the
l.h.s. being subdominant). One can be more precise and rescale $\xi=x/p$. The new probability
$\tilde \omega(p,\xi)=\omega(p,x)$ is solution of the differential equation
\begin{equation}
p^2\, \frac{\partial \tilde{\omega}(p,\xi)}{\partial p}  = \tilde v(p,\xi) \frac{\partial \tilde{\omega}(p,\xi)}{\partial \xi} + \frac{\kappa}{2} \frac{\partial^2 \tilde{\omega}(p,\xi)}{\partial \xi^2}\label{fp},
\end{equation}
where $\tilde v(p,\xi)=p (v(p,x)+\xi)$. Inserting the single shock profile one finds
$\tilde v_s(p,\xi) = \pi \tanh(\pi \xi/2)$, i.e. a stationary velocity field. Hence we may convince ourselves that the
stationary solution of (\ref{fp}) is $\partial_x \omega_{stat}=\lambda_{d}$, i.e. the
convergence to the dipolar result \eqref{eqn:dipolarepd} includes also (non exponential) subdominant terms in the scaling region $x \sim p$. Interesting exact solutions of (\ref{fp}) for all $p$ are easily obtained from a Laplace transform in $\tau=1/p$ or using methods as in \cite{gorodtsov:99}, but do not obviously relate to SLE.

\subsection{Results for $\kappa=2$ and the winding of loop-erased random walks}
\label{sec:kappa2}

\begin{figure}
  \begin{center}
  \begin{tabular}{cp{1cm}c}
    \includegraphics[height=5.5cm]{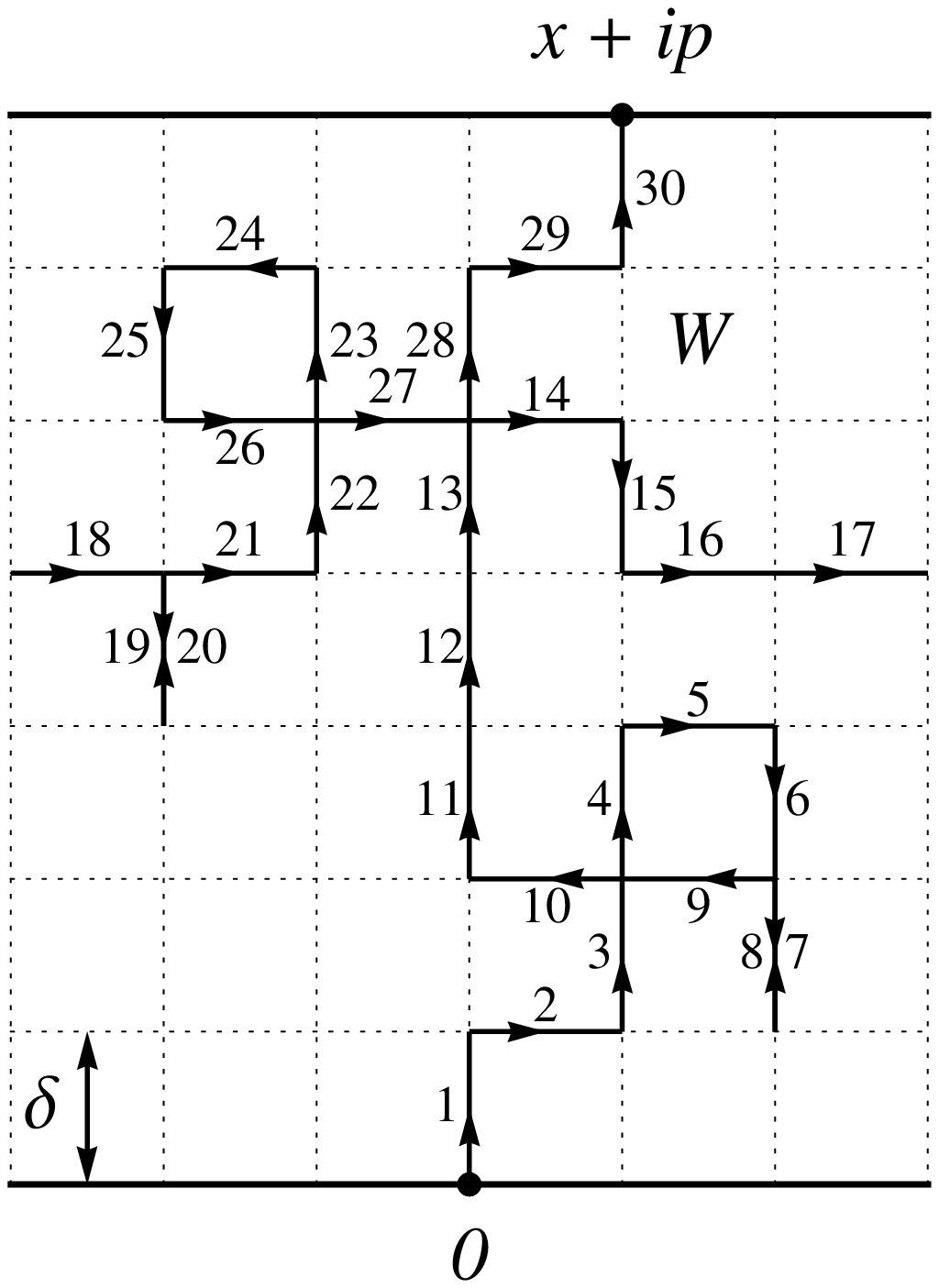}&&
    \includegraphics[height=5.5cm]{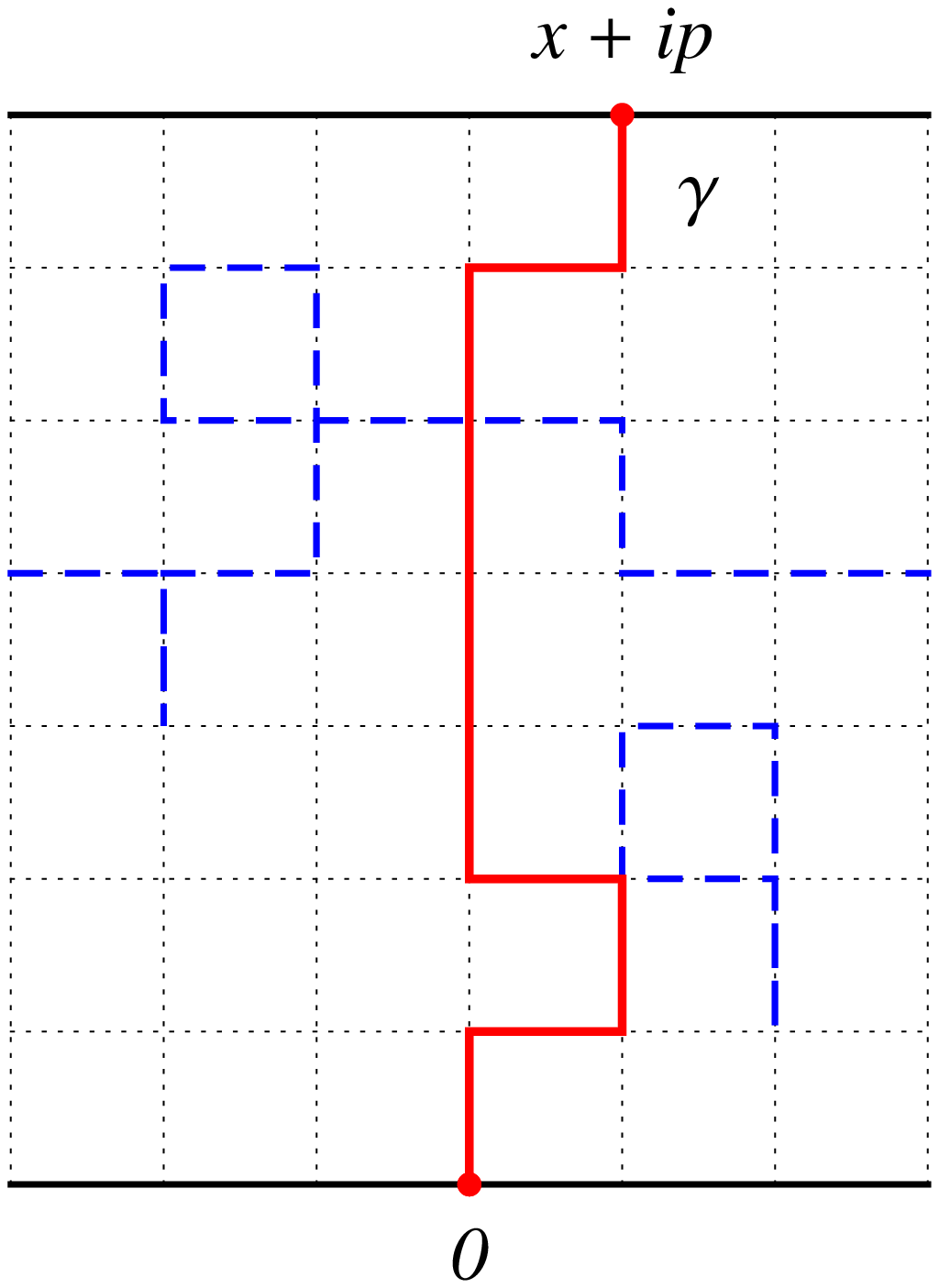}\\
    (a) && (b)
  \end{tabular}
  \end{center}
  \caption{Illustration of loop-erasure on the cylinder for a square lattice. (a) A Brownian excursion from $z=0$ to the upper boundary $y=p$. (b) Loop-erasure of the same walk. Loops like $(14,\,15,\dots,\,27)$, winding around the cylinder for several times, may be erased.}
  \label{fig:lerw}
\end{figure}

The Cole-Hopf transformation \eqref{eqn:colehopf} will lead to interesting simplifications allowing to treat explicitly $\kappa=2$. This case corresponds to the scaling limit of the LERW on the cylinder. First we recall its definition on the lattice \cite{lawler:06}. Consider a rectangular lattice domain of lattice mesh $\delta$ embedded into the cylinder domain $\mathbb{T}_p$. Start a simple random walk $\mathcal{W}$ from $0$ to $\text{Re}\, z = y = p$ on with absorbing boundary conditions for $y=0$ (see figure \ref{fig:lerw}a). Chronological loop erasure from $\mathcal{W}$ yields a simple path $\gamma$ on the lattice domain which is the LERW (see figure \ref{fig:lerw}b). Notice that the endpoints of $\mathcal{W}$ and $\gamma$ always coincide. However, the random walk may wind several times around the cylinder whereas its loop erasure does not. This makes it a non-trivial exercise to compute the winding properties of the latter (i.e. its endpoint distribution on the covering space). As the lattice mesh $\delta$ goes to $0$, the paths $\mathcal{W}$ converge to two-dimensional Brownian excursions on $\mathbb{T}_p$ whereas the paths $\gamma$ converge to SLE$_2$ on $\mathbb{T}_p$ \cite{zhan:06_2}.

Since the endpoints of $\mathcal{W}$ and $\gamma$ coincide on the cylinder $\mathbb{T}_p$, we may find the endpoint distribution for SLE$_2$ by a simple computation on Brownian excursion on the cylinder, starting from $z=0$ and stopped as soon as it hits altitude $y=p$ (see \ref{sec:abc}). The result is
\begin{equation}
  \Lambda_2(p,x) = \frac{\pi}{4p}\sum_{n=-\infty}^\infty \frac{1}{\left(\cosh\left(\displaystyle\frac{\pi(x+2\pi n)}{2p}\right)\right)^2}.
  \label{eqn:brownian}
\end{equation}
An equivalent result for random walks on a finite cylindrical lattice domain is presented in appendix \ref{sec:lattice}.
Indeed, $\Lambda_2(p,x)$ solves \eqref{eqn:pde} for $\kappa=2$ with periodic boundary conditions and may be related to the drift function $v(p,x)$ via $\Lambda_2(p,x) = (p \,\partial v(p,x)/\partial x+1)/2\pi$  \cite{zhan:06}. Moreover, it corresponds to the periodised endpoint distribution \eqref{eqn:dipolarepd} for dipolar SLE with $\kappa = 2$.

\begin{figure}
\begin{center}
\includegraphics[height=4.5cm]{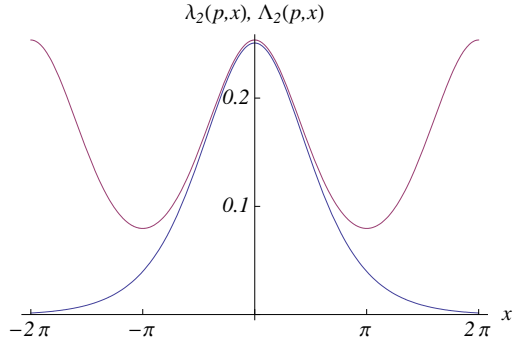}
\end{center}
\caption{ The endpoint distributions $\Lambda_2(p,x)$ and $\lambda_2(p,x)$ for SLE$_2$ in the case $p=\pi$.}
\label{fig:distributions}
\end{figure}

In order to compute the winding probabilities and to find the distribution on the covering space $\lambda(p,x)$, we solve \eqref{eqn:proba} with boundary conditions as indicated above.
The idea is to write $\omega(p,x) = \alpha(p,x) \psi(p,x)$
and impose that terms proportional to $\partial\psi(p,x)/\partial x$ 
cancel, which, for arbitrary $\kappa$, yields $\alpha(p,x) \propto \epsilon(p,x)^{-2/\kappa}$ \cite{gorodtsov:99}. It turns out that
the function $\psi(p,x)$ is solution to
\begin{equation}
  \frac{\partial \psi(p,x)}{\partial p} =\left(\frac{2-\kappa}{2\kappa}\right)\frac{\partial v(p,x)}{\partial x}\psi(p,x)+\frac{\kappa}{2}\frac{\partial^2\psi(p,x)}{\partial x^2}.
  \label{eqn:psi}
\end{equation}
Hence for $\kappa=2$ we have a simple diffusion equation $\partial \psi(p,x)/\partial p = \partial^2\psi(p,x)/\partial x^2$ with initial condition
\begin{equation}
  \psi(0,x)=\epsilon(0,x)\omega(0,x)=2 \pi \sum_{n=-\infty}^{\infty}\delta(x-\pi -2\pi
  n)\Theta(x).
\end{equation}
In this case the solution therefore is immediate:
\begin{align}
 \psi(p,x)&=\int_{-\infty}^{\infty}\frac{\diff y}{\sqrt{4\pi
  p}}\,e^{-(x-y)^2/4p}\, \psi(0,y)
  =\sqrt{\frac{\pi}{p}}  \sum_{n=0}^{\infty}e^{-(x-\pi(2n+1))^2/4p}.
\end{align}
Thus, for $\kappa = 2$ the endpoint probability reads
\begin{align}
\omega_2(p,x)&= \frac{\sum_{n=0}^{\infty}
  \exp\left(-(x-\pi(2n+1))^2/4p\right)}
  {\sum_{n=-\infty}^{\infty}
  \exp\left(-(x-\pi(2n+1))^2/4p\right)}=\frac{\sum_{n=0}^{\infty}
  \exp\left(\pi n( x -\pi(n+1))/p\right)}
  {\sum_{n=-\infty}^{\infty}
  \exp\left(\pi n( x -\pi(n+1))/p\right)}.
  \label{eqn:omega}
\end{align}
Note that $\omega(p,x)=\psi(p,x)/(\psi(p,x)+\psi(p,-x))$. The endpoint distribution is given by
\begin{equation}
\lambda_2(p,x)= \frac{\sum_{n,m=0}^{\infty}(m{+}n{+}1)
  \exp\left(\displaystyle -\frac{(x{-}\pi{-}2\pi n)^2+(x{+}\pi{+}2\pi m)^2}{4p}\right)}
  {(p\,\epsilon(p,x)/\pi)^2}
  \label{eqn:lambda}
\end{equation}
This result is exact and allows to analyse for $\kappa=2$ how the
SLE trace wraps around the cylinder.  An illustration of the distributions is given in figure \ref{fig:distributions}.
Some remarks are at order. First of all, we have checked through tedious calculation that
\eqref{eqn:lambda} inserted in \eqref{eqn:periodisation} correctly reproduces \eqref{eqn:brownian}. Second, the very particular form of \eqref{eqn:omega}, given by a ratio of propagators for one-dimensional Brownian motion, suggests that it might be obtained as a conditional probability for a simple one-dimensional process. Indeed, such a relation can be established (see section \ref{sec:probarg}). Moreover, notice an interesting relation to (boundary) conformal field theory. Setting $q=\exp(-2\pi^2/p)$, we see that the different terms in \eqref{eqn:omega} contain $q^{h_{1,n+1}}$ where $h_{1,n} = \dim \psi_{1,n} = n(n-1)/2$ are the scaling dimensions of the operators $\psi_{1,n}$, creating $n-1$ curves at the boundary, for the $O(N=-2)$ model \cite{cardy:06}. Finally, one may hope that the simplifications in the case $\kappa=2$ for the endpoint probabilities may be extended to bulk probabilities of the cylinder and the covering space, a case which we treat (partially) in appendix \ref{sec:bulk}.

The closed forms \eqref{eqn:omega} and \eqref{eqn:lambda} converge well in the dipolar limit $p\to 0^+$, however are not suitable for computations in the radial limit $p\to +\infty$. Therefore, we study the moment generating function $\langle e^{-\mu x}\rangle$ here below, and shall give an exact expression adapted to both limits. To this end, it will be useful to consider $\omega_2(p,z) = \psi(p,z)/\epsilon(p,z)$ for complex $z$. From \eqref{eqn:omega} we see that it is periodic with period $2ip$. Moreover, since the denominator can be written as an infinite product \cite{abramowitz:70}
\begin{equation}
  \epsilon(p,z) \propto \prod_{k=1}^{\infty}\left(1-e^{iz-(2k-1)p}\right)\left(1-e^{-iz-(2k-1)p}\right)
\end{equation}
(up to a $p$-dependent prefactor), we conclude that the function $\omega_2(p,z)$ has simple poles in the complex plane for $z_{k,\ell} = (2k-1)ip + 2\pi \ell,\,k,\ell\in \mathbb{Z}$. In order to compute the generating function, we write
\begin{equation}
   \langle e^{-\mu x}\rangle = \mu \int_{-\infty}^\infty\diff x\,e^{-\mu x}\omega_2(p,x),\qquad \mu > 0.
\end{equation}
Deformation of the integration contour in the complex plane from $\mathbb{R}$ to $\mathbb{R}+2ip$ leads to two distinct contributions. First, the integration along $\mathbb{R}+2ip$ simply yields $e^{-2ip\mu}\langle e^{-\mu x}\rangle$ because of the periodicity of $\omega_2(p,z)$. Second, we must take into account the simple poles at $z_\ell=z_{1,\ell}=ip+2\pi \ell,\,\ell\in\mathbb{Z}$ and compute the corresponding residues. Solving for $\langle e^{-\mu x}\rangle$, we find
\begin{align}
  \langle e^{-\mu x}\rangle &= \frac{2\pi i\mu}{1-e^{-2ip\mu}}\sum_{l\in \mathbb{Z}}
    \mathop{\mathrm{res}}_{z=z_{l}}e^{-\mu z}\omega_2(p,z)\nonumber\\&=\frac{\pi\mu}{\sin p\mu}\sum_{l\in \mathbb{Z}}{e^{-2\pi \mu \ell} \psi(p,ip+2\pi \ell)}\left(\frac{\partial\epsilon(p,ip+2\pi\ell)}{\partial z}\right)^{-1}
\end{align}
The sum can explicitly be computed because the derivative of $\epsilon(p,z)$ at $z=z_\ell$ does not depend on $\ell$ by periodicity. After a little algebra we obtain
\begin{align}
  \langle e^{-\mu x}\rangle= \frac{p\mu}{\sin p\mu}\,\frac{1}{\sinh \pi \mu}\,
  \frac{\sum_{\ell=0}^\infty(-1)^\ell\sinh ((2\ell+1)\pi \mu)e^{-\pi^2\ell(\ell+1)/p}}{\sum_{\ell=0}^\infty(-1)^\ell(2\ell +1)e^{-\pi^2\ell(\ell+1)/p}}.
  \label{eqn:smallp}
\end{align}
For large $p$ it is convient to transform this expression into a dual series
\begin{equation}
\langle e^{-\mu x}\rangle = \frac{e^{p\mu^2}\pi \mu}{\sinh \pi \mu}\,\frac{1}{\sin p\mu}\,
  \frac{\sum_{\ell=0}^\infty(-1)^\ell\sin ((2\ell+1)p \mu)e^{-p\ell(\ell+1)}}{\sum_{\ell=0}^\infty(-1)^\ell(2\ell +1)e^{-p\ell(\ell+1)}}
  \label{eqn:largep}
\end{equation}
by using identities for Jacobi theta functions \cite{abramowitz:70} or Poisson's summation formula.
Let us now study the radial and dipolar limit of our result.

\medskip

\textit{Limit $p\to+\infty$.} In the radial limit, we may write \eqref{eqn:largep} as
\begin{equation}
   \langle e^{-\mu x}\rangle
  =\frac{\pi \mu\,e^{p\mu^2}}{\sinh\pi \mu}\left(1{+}2e^{-2p}(1{-}\cos 2p\mu)\right)+O(e^{-4p}).
\end{equation}
In fact, it is possible to recover this result by approximation of $\epsilon(p,x)$ with its series expansion \eqref{eqn:seriesepsilon} and Gaussian integration. The result turns out to be finite result for
all $\mu$ since the probability decays as a Gaussian at large $x$, i.e. $1-\omega_2(p,x) = \omega_2(p,-x) \sim \exp(-(x+\pi)^2/4p)$  as ${x \to + \infty}$, up to a prefactor periodic in $x$. 
We take the logarithm and find a cumulant expansion
\begin{align}
 \log\langle e^{-\mu x}\rangle &=\sum_{k=1}^\infty(-1)^k\langle x^k\rangle_c\,\frac{\mu^k}{k!}
 = p\mu^2+\log\left(\frac{\pi \mu}{\sinh\pi \mu}\right)+2e^{-2p}(1-\cos 2p\mu)+O\left(e^{-4p}\right).
\end{align}
The result shows that, as announced above, we do not recover a Gaussian distribution in the limit $p\to+\infty$, in contrast to the simple argument from radial SLE. It rather provides all the constants
$X_{2n,\kappa=2}$ from the Taylor series expansion of the function $\log [\pi\mu/(\sinh\pi\mu)]$ with respect to $\mu$. For instance, we find
\begin{align}
\langle x^2\rangle_c = 2 p -\pi^2/3 + 8 p^2 e^{-2 p} +O\left(e^{-4p}\right)\\
\langle x^4\rangle_c= 2\pi^4/15 - 32 p^4 e^{-2 p} +O\left(e^{-4p}\right)
\end{align}
up to corrections of the form $p^n e^{-4 p}$.

\medskip

\textit{Limit $p\to 0^+$.} In the dipolar limit, we expand the cumulant generating function, using \eqref{eqn:smallp},
as
\begin{equation}
  \log\langle e^{-\mu x}\rangle = \log\left(\frac{p\mu}{\sin p\mu}\right) - 4 \sinh^2(\pi \mu) e^{-2\pi^2/p}.
\end{equation}
Hence the first two non-vanishing cumulants are given by
\begin{align}
  \langle x^2\rangle_c = p^2/3 - 8 \pi^2 e^{-2\pi^2/p} +O\left(e^{-4\pi^2/p}\right),\\
\langle x^4\rangle_c= 2p^4/15 - 32 \pi^4 e^{-2\pi^2/ p} +O\left(e^{-4\pi^2/p}\right).
\end{align}
Since the first corrections to the dipolar result at small $p$ originate from erasure of a single loop wrapping around the cylinder, it seems plausible that the probability for erasing such a loop behaves like $\exp(-2\pi^2/p)$ in the limit $p\to 0^+$. This factor has indeed the same magnitude as the probability that planar Brownian motion on $\mathbb{T}_p$ winds by at least $2 \pi$ around the cylinder (we recall the winding distribution for Brownian motion in appendix \ref{sec:abc}).

Notice that the dipolar limit presents a subtlety, which does not allow to conclude immediately from expansion of the probability $\omega_2(p,x)$. For any fixed $x$ such that $- \pi <x < \pi$ we find estimate as $p\to 0^+$ (a neighbourhood of size $\sim p$ of the points $\pm \pi$ must be excluded):
\begin{equation}
\omega_2(p,x) = \frac{1}{1+e^{-\pi x/p}} + \tanh\left(\frac{\pi x}{2 p}\right) e^{-2\pi^2/p} 
+ o(e^{-2\pi^2/p}). \label{limdip}
\end{equation}
Up to exponential corrections, and uniformly in the interval $]- \pi,\pi[$, we recover the dipolar result \eqref{eqn:dipolarepd} for $\kappa=2$. However, the approximation is not uniform beyond this interval what leads to complications when computing moments by using this approximation.

\subsection{Probabilistic argument for $\kappa=2$}
\label{sec:probarg}

The simplifications for SLE$_2$ seem surprising at first sight. Here we shall try to understand the underlying mechanism by giving a probabilistic argument for the simple solution of \eqref{eqn:proba} by \eqref{eqn:omega} in the case $\kappa=2$.

Let us consider the stochastic process $X_t = g_t(x+ip)-i(p-t)-\sqrt{\kappa} B_t$ with initial condition $X_0=x$. It describes the motion of a given point $x+ip$ on the upper boundary of the cylinder under the flow $g_t(z)$ and is solution of the stochastic differential equation
\begin{equation}
  \diff X_t = v(p-t,X_t)\diff t +\sqrt{\kappa} \diff B_t.
\end{equation}
In particular, it has been shown that $\lim_{t\to p^-}X_t = (2k+1)\pi$ with $k\in \mathbb{Z}$ (almost surely) \cite{zhan:06}. We may relate the probability that the SLE trace hits the upper boundary of the covering space within $(-\infty,x]$ to $X_t$ via
\begin{align}
\omega(p,x)&=\mathsf{P}[\text{Re}\,\gamma_p \in (-\infty, x]]
=\mathsf{P}[\text{Re}\,g_t(\gamma_p) \in (-\infty, \text{Re}\,g_t(x+ip)]]\nonumber\\
 &\mathop{=}_{t\to p}\mathsf{P}[\sqrt{\kappa}\,B_p \in (-\infty, \text{Re}\,g_p(x+ip)]]= \mathsf{P}_x[X_{p}\geq 0],
 \label{eqn:confinv}
\end{align}
where $\mathsf{P}_x$ denotes ``probability'', taking into account that the process starts from $x$. Along these lines we have used the conformal invariance of the SLE measure, leading to invariance of probabilities under conformal transport.

We now show that the right hand side of \eqref{eqn:confinv} may be computed easily by establishing a relation to conditioned Brownian motion.
Consider the process $x_t=\sqrt{\kappa}B_t$ of simple Brownian motion with diffusion constant $\kappa$ on the interval $I=[-\pi,\pi]$, identifying its endpoints, which starts from $x_0=x\in I$. We introduce the propagator via $P(y,t;x)\diff y = \mathsf{P}_{x}[x_t\in [y,y+\diff y]]$. It is solution of the diffusion equation $\partial P(y,t;x)/\partial t= (\kappa/2)\partial^2 P(y,t;x)/\partial x^2$, and explicitly given by
\begin{equation}
  P(y,t;x)=\frac{1}{\sqrt{2\pi \kappa t}}\sum_{n=-\infty}^\infty\exp\left(-\frac{(y-x-2\pi n)^2}{2\kappa t}\right).
\end{equation}
Let us condition $x_t$ to arrive at $x_p=\pi \,(\text{mod}\,2\pi)$ for some given time $p>0$. Using elementary facts about conditional probabilities we see that the new process has a propagator $Q(y,t;x)$ defined via
\begin{align}
  Q(y,t;x)\diff y &= \mathsf{P}_{x}[x_t\in [y,y+\diff y]\mathop{|}x_p=\pi\,(\text{mod}\,2\pi)]\nonumber\\
  &= \frac{P(\pi,p-t;y)}{P(\pi,p;x)}\,P(y,t;x)\diff y,\quad\text{for }0<t<p.
\end{align}
The conditioning therefore leads to a drift \cite{doob:83} that can be read off from the diffusion equation for $Q(y,t;x)$:
\begin{equation}
\frac{\partial Q(y,t;x)}{\partial t} = -\kappa\frac{\partial}{\partial y}\left(\left(\frac{\partial}{\partial y} \ln P(\pi,p-t;y)\right)Q(y,t;x)\right)+\frac{\kappa}{2}\frac{\partial^2 Q(y,t;x)}{\partial y^2}
\end{equation}
Specialising to $\kappa=2$, we conclude that the conditioned process is solution of the stochastic differential equation
\begin{align}
  \diff x_t &= 2\left(\frac{\partial}{\partial x} \ln P(\pi,p-t;x_t)\right)\diff t+\sqrt{2}\,\diff B_t \nonumber \\
  &= v(p-t,x_t)\diff t + \sqrt{2}\,\diff B_t 
\end{align}
Hence we see that in the case $\kappa=2$ the motion $X_t$ induced by the Loewner mapping $g_t(z)$ on the upper boundary of the cylinder is the same as Brownian motion $x_t$ on a circle, starting from $x$ and conditioned to visit $\pi \,(\mathop{\text{mod}}\,2\pi)$ at time $t=p$. However, it is a simple exercise to compute the statistics of the winding number of the latter. Again, we shall say that $x_t$ has winding number $N=n$ if its equivalent on the covering space $\mathbb{R}$ of $I$ arrives at $x+2\pi n$ at time $t$. From simple conditioning, we have
\begin{equation}
\mathsf{P}_x[N=n\mathop{|} x_p = \pi\,(\text{mod}\,2\pi)]=\frac{\exp\left(-{(\pi-x+ 2\pi n)^2}/{2\kappa p}\right)
}{\sum_{n=-\infty}^\infty\exp\left(-{(\pi-x+2\pi n)^2}/{2\kappa p}\right)}
\end{equation}
Summation over $n$ from $0$ to $\infty$ yields the announced result for $\kappa=2$
\begin{equation}
\omega(p,x)=\mathsf{P}_x[X_p\geq 0]=\frac{\sum_{n=0}^{\infty}
  \exp\left({-(x-\pi(2n+1))^2/4p}\right)}
  {\sum_{n=-\infty}^{\infty}
  \exp\left({-(x-\pi(2n+1))^2/4p}\right)}
\end{equation}
Thus we have shown that we may indeed reinterpret the winding of SLE$_2$ in terms of the winding of a conditioned one-dimensional Brownian motion. In fact, this is the profound reason explaining why the preceding transformation leads to simplifications for $\kappa=2$.

\subsection{The winding of loop-erased random walks with fixed endpoints}
In this section we consider the case of a LERW with fixed endpoints from $0$ to $x+ip$, $-\pi \leq x \leq \pi$ in the cylinder geometry. On the covering space, it is thus allowed to exit at $x+2\pi N+ip$ with arbitrary winding number $N\in \mathbb{Z}$. We shall be interested in the law of $N$ which may be obtained by conditioning LERWs to exit at the given boundary points.

Let us ask for the probability that the LERW has made $N=n$ windings if we condition its trace to exit on a subinterval $[a,b]+ip$, $-\pi \leq a,b\leq \pi $ on the upper boundary. It is given by
\begin{equation}
 \mathsf{P}[N=n \mathop{|} \text{Re}\,\gamma_p \in [a,b]]=\frac{\mathsf{P}[N=n \cap \text{Re}\,\gamma_p \in [a,b]]}{\mathsf{P}[\text{Re}\,\gamma_p \in [a,b]]}
=\frac{\int_a^b\diff x\,\lambda_2(p,x+2\pi n)}{\int_a^b\diff x\,\Lambda_2(p,x)},
\end{equation}
where $\Lambda_2(p,x)$ and $\lambda_2(p,x)$ are the distributions defined in (\ref{eqn:brownian},\ref{eqn:lambda}).
Taking the limit $a,b\to x$ amounts to forcing the LERW to exit at $x+ip$. It has been noted previously that (at least for simply-connected domains) this procedure of conditioning leads to chordal SLE$_2$ from $0$ to $x+ip$ \cite{bauer:07, bauer:08, lawler:08}. Supposing that this property holds for doubly-connected domains as well, we obtain the statistics of the winding number of SLE$_2$ with fixed endpoints $0$ and $x+ip$ on the boundary of $\mathbb{T}_p$:
\begin{equation}
  \mathsf{P}_{0\to x+ip}[N=n]= \frac{\lambda_2(p,x+2\pi n)}{\Lambda_2(p,x)}=\frac{\lambda_2(p,x+2\pi n)}{\displaystyle\sum_{k=-\infty}^\infty\lambda_2(p,x+2\pi k)}
  \label{eqn:windprob}
\end{equation}
where $-\pi <x<\pi$
This result is non-trivial (at least for us), since there does not seem to exist an obvious way to recover this probability law from considerations of underlying random walks/Brownian motions.

As an application of \eqref{eqn:windprob} we compute the generating function for the moments of $N$. For $\mu>0$ let us write
\begin{align}
 \left\langle e^{-\mu N}\right\rangle& = \frac{1}{\Lambda_2(p,x)}\sum_{n=-\infty}^\infty e^{-\mu n}\lambda_2(p,x+2\pi n)\nonumber\\
 & = \frac{1}{\Lambda_2(p,x)}\frac{\partial}{\partial x}\left(\frac{1}{\epsilon(p,x)}\sum_{n=-\infty}^\infty e^{-\mu n}\psi(p,x+2\pi n) \right)\nonumber\\
 &= \frac{1}{\Lambda_2(p,x)}\frac{ e^{p\mu^2/4\pi^2}}{2\,\sinh(\mu/2)}\frac{\partial}{\partial x}\left(\frac{e^{\mu x/2\pi}\epsilon(p,x+p\mu/\pi)}{\epsilon(p,x)}\right)
\end{align}
Strictly speaking, the calculation is valid for $\mu >0$ but the result extends to negative $\mu$, too. The result may be used in order to study the winding behaviour in the radial and dipolar limit.

\medskip

\textit{Radial limit.} For $p\to +\infty$, we find a generating function for the cumulants given by
\begin{equation}
  \log \left\langle e^{-\mu N}\right\rangle = \frac{\mu x}{2\pi} + \frac{p\mu^2}{4\pi^2}+\log\left(\frac{\mu/2}{\sinh (\mu/2)}\right)+O(e^{-p}),
\end{equation}
where we have used in that limit $\Lambda_2(p,x) \to 1/(2 \pi)$. 
As in section \ref{sec:kappa2} the cumulants $\langle N^n\rangle_c$ with $n>2$ converge to constants which can be obtained from the Taylor series expansion of the function $\log [\mu/(2\sinh(\mu/2))]$ with respect to $\mu$. Corrections to this limit are exponentially small $\propto e^{-p}$.

\medskip

\textit{Dipolar limit.} Having in mind the discussion in section \ref{sec:kappa2}, we expect the probability distribution of $N$ to be concentrated at $N=0$ as $p\to 0^+$ (independently of the value taken by $|x| < \pi$), and the probabilities for events occurring with $N\neq 0$ to be exponentially small $\propto e^{-2\pi^2/p}$. By means of explicit computation we find
\begin{equation}
 \left\langle e^{-\mu N}\right\rangle = 1 + 4\sinh\left(\frac{\mu}{2}\right)\left(2\sinh\left(\frac{2\pi x+\mu p}{2p}\right)+\sinh\left(\frac{4\pi x+\mu p}{2p}\right)\right)e^{-2\pi^2/p}
\end{equation}
up to corrections of the order of $O(e^{-4\pi^2/p})$, what confirms the preceding discussion. 

\section{Endpoint probabilities for $\mathbf{\kappa=4}$ on the cylinder}
\label{sec:kappa4}

It is surprising that we obtain the endpoint distribution on the cylinder $\kappa=2$ by periodisation of the equivalent dipolar result \eqref{eqn:dipolarepd}.
For $\kappa=4$ we have applied the same idea to look for a solution in
the form of a $2\pi$-periodised result found in dipolar
SLE$_4$. These two values seem to be the only ones for which the periodisation yields exact results. The case $\kappa=4$ is related to the massless free field theory with central charge $c=1$. In fact, SLE$_4$ curves arise as zero lines of the field $\varphi(z,\overline{z})$ emerging from discontinuities in Dirichlet boundary conditions. Moreover, SLE$_4$ arises in the scaling limit of several lattice models such as the harmonic explorer and domino tilings \cite{schramm:06,kenyon:00}. A variant of the harmonic explorer leading SLE$_4$ in doubly-connected domains as discussed in this article was suggested in \cite{zhan:06}.

In this section, we shall show that the periodic endpoint distribution 
\begin{equation}
 \Lambda_4(p,x)=\frac{1}{2
  p}\sum_{n=-\infty}^{\infty}\frac{1}{\cosh\displaystyle\left[
  \frac{\pi (x+2\pi n)}{2p}\right]}
  \label{eqn:reflbrownian}
\end{equation}
is solution of \eqref{eqn:pde} with $\kappa=4$.
Let us notice, that \eqref{eqn:reflbrownian} corresponds to the endpoint distribution of two-dimensional Brownian motion on the cylinder with reflecting boundary conditions for $y=0$, stopped for the first time when it reaches altitude $y=p$ (see appendix \ref{sec:mbc}).

We analytically continue $\Lambda_4(p,z)$ to complex
$z\in \mathbb{C}$ and introduce the function
\begin{equation}
  g(p,z) = \frac{\partial\Lambda_4(p,z)}{\partial p}- \frac{\partial }{\partial
  z}(v(p,z) \Lambda_4(p,z))-2\frac{\partial^2 \Lambda_4(p,z)}{\partial
  z^2}
  \label{eqn:g}
\end{equation}
with $v(z,p)$ as defined previously. If $\Lambda_4(p,z)$ is a solution of 
(\ref{eqn:pde}) with $\kappa=4$ then $g(p,z)$ must vanish for
$z\in \mathbb{R}$. We shall even show that $g(p,z)$ vanishes for
$\Lambda_4(p,z)$ defined in \eqref{eqn:reflbrownian} for all $z\in
\mathbb{C}$.

$\Lambda_4(p,z)$ is elliptic
with periods $\omega_1 = 2\pi$ and $\omega_2 = 4p i$ and
antiperiodic with respect to the half-period $\omega_2/2 = 2 p i$.
Thus, $g(p,z)$ is elliptic for all $p$ with the same periods
$\omega_1$ and $\omega_2$ (as $z \to z+4 p i$, the additional term
arising from the quasi-periodicity of $v(p,z)$ is compensated by the
derivative of $\lambda$ with respect to $p$). We may therefore
restrict our analysis to the rectangle defined via $-\pi < \text{Re}\,z<\pi$ and $0<\text{Im}\,z < 4p$. The poles of $\Lambda_4(p,z)$, $v(p,z)$ and their derivatives are located
at $z=ip$ and $z=3ip$. The principal parts of the Laurent series expansions at $z=ip$ for the different terms in (\ref{eqn:g}) read
\begin{align}
  &\frac{\partial \Lambda_4(p,z)}{\partial p} =
  \frac{1}{\pi (z-ip)^2}+O(1),\,
  \frac{\partial^2 \Lambda_4(p,z)}{\partial z^2} =
  -\frac{2i}{\pi(z-ip)^3}+O(1) \nonumber\\
  &\frac{\partial (v(p,z)\Lambda_4(p,z))}{\partial z} =
  \frac{1}{\pi(z-ip)^2}+\frac{4i}{\pi(z-ip)^3}+O(1)
\end{align}
Insertion into therefore shows that all divergent terms cancel out so that $g(p,z)$ has a removable singularity at $z=ip$. The Laurent expansion around $z=3ip$ leads to
the same result. In fact, we have
\begin{align}
  &\frac{\partial \Lambda_4(p,z)}{\partial p} =
  -\frac{3}{\pi(z-3ip)^2}+O(1),
  \frac{\partial^2 \Lambda_4(p,z)}{\partial z^2} =
  \frac{2i}{\pi(z-3ip)^3}+O(1) \nonumber\\
  &\frac{\partial (v(p,z) \Lambda_4(p,z))}{\partial z} =
  -\frac{3}{\pi(z-3ip)^2}-\frac{4i}{\pi(z-3ip)^3}+O(1)
\end{align}
Therefore $g(p,z)$ has a removable singularity at $z=3pi$, too.
Hence it must be a constant $g(p,z) \equiv A(p)$ what follows from Liouville's theorem \cite{ahlfors:79}. However,
because of $g(p,z) = -g(p,z+2p i)$ we find $A(p) = 0$ for all $p$,
so that $g(p,z) \equiv 0$ which is equivalent to say that for
$\kappa=4$ the given form for $\Lambda_4(p,x)$ is the correct
probability distribution function.

Let us note that \eqref{eqn:reflbrownian} may be written in terms of the Burgers potential $\epsilon(p,x)$.
\begin{equation}
  \Lambda_4(p,x) = \mathcal{N}(p)\,\frac{\epsilon(p,x-\pi)}{\epsilon(p,x)}.
\end{equation}
The normalisation factor $\mathcal{N}(p)$ may be found in a similar way as \eqref{eqn:smallp} and \eqref{eqn:largep}, and is given by
\begin{equation}
  \mathcal{N}(p) =\frac{1}{\pi \epsilon(p,ip-\pi)}\frac{\partial\epsilon(p,ip)}{\partial z }= \frac{1}{\pi}\frac{\sum_{n\in \mathbb{Z}} (-1)^n n\, e^{-n(n+1)p}}{\sum_{n\in \mathbb{Z} } e^{-n(n+1)p}}.
\end{equation}
However, despite this very suggestive form of $\Lambda_4(p,x)$ as a ratio of two solutions to the simple diffusion equation we have not found algebraic simplifications in order to solve the partial differential equations for $\kappa=4$ with general boundary conditions.

\section{Conclusion}

In this paper, we have studied winding properties of loop-erased random walks around a finite cylinder by means of stochastic Loewner evolutions. Relating the computation of the endpoint distribution $\lambda(p,x)$ of SLE$_\kappa$ on a cylinder to a problem of diffusion-advection of a passive scalar in a Burgers flow, we were able to explicitly determine $\lambda_2(p,x)$ in the case $\kappa=2$. The behaviour in the limit of very thin and very large cylinders was studied and we pointed out a non-Gaussian behaviour of the winding properties for long cylinders. Furthermore, conditioning the loop-erased random walks to exit via a given boundary point, we were able to compute the probability distribution of the winding number for walks with fixed endpoints. We have shown that the somewhat surprising simplifications for $\kappa=2$ may be related to the winding of conditioned one-dimensional Brownian motion on a circle. Moreover, we have determined cylinder endpoint distribution in the case $\kappa=4$. A relation to reflected Brownian motion was pointed out. As for $\kappa=2$, it is related to the periodisation of the dipolar endpoint distribution. However, these seem to be the only values for $\kappa$ for which this property holds. It remains to see whether closed results can also be obtained for other values of $\kappa$. We hope that our results can be useful to test recent conjectures \cite{alan} about SLE properties of interfaces in numerical studies in a cylinder geometry.

\section*{Acknowledgements}
We would like to thank Michel Bauer, Denis Bernard and Alan Middleton for very useful discussions.
PLD acknowledges support from ARN BLAN05-0099-01. CH benefits from financial support from the French Minist\`ere de l'Education et de la Recherche.

\appendix

\section{Endpoint distributions for two-dimensional Brownian motion}
In this appendix we compute the endpoint distribution for planar Brownian motion $z_t=x_t+iy_t$  on the cylinder, starting from $z=0$ and stopped as soon as it reaches the altitude $y=p$. For simplicity, we first consider the geometry of an infinite strip $\{z\in \mathbb{C}\mathop{|}0<\text{Im}\,z<p\}$ and then periodise along the real axis in order to obtain the results on the cylinder $\mathbb{T}_p$. The diffusion in $x-$direction is unconstrained and has a propagator $\mathsf{P}_{x_0=0}[x_t\in[x,x+\diff x]] =\exp (-x^2/2t)\, \diff x/\sqrt{2\pi t}$. For the motion in $y-$direction, we consider absorbing and mixed boundary conditions.

\subsection{Absorbing boundary conditions}
\label{sec:abc}
Consider the motion in $y$-direction, starting from $y_0=\epsilon >0$. We shall impose absorbing boundary conditions at $y=0$ and $y=p$, and condition the process to exit at $y=p$.
The propagator $P_A(y,t)$ defined via $P_A(y,t)\diff y= \mathsf{P}_{y_0=\epsilon}[y_t\in [y,y+\diff y]]$ is solution of the diffusion equation $\partial P_A(y,t)/\partial t =1/2\, \partial^2 P_A(y,t)/\partial y^2$ and given by
\begin{equation}
  P_A(y,t) = \frac{2}{p}\sum_{n=1}^\infty \sin\left(\frac{n\pi y}{p}\right)\sin\left(\frac{n\pi \epsilon}{p}\right)\exp\left(-\frac{n^2\pi^2t}{2p^2}\right)
\end{equation}

We obtain the exit-time distribution $f_A(t)$ at $y=p$ from the probability current at this point. However, if we condition the diffusion to exit at $y=p$, we furthermore must divide this current by the exit probability $\mathsf{P}_{y_0=\epsilon}[\text{exit at } y=p] = \epsilon/p$. Hence
\begin{align}
f_A(t) &= -\left.\frac{p}{2\epsilon}\frac{\partial P(y,t; \epsilon)}{\partial y}\right|_{y=p}= \frac{\pi}{p\epsilon}\sum_{n=1}^\infty (-1)^{n+1} n\sin\left(\frac{n\pi \epsilon}{p}\right)\exp\left(-\frac{\pi^2 n^2 t}{2p^2}\right)\nonumber\\
 &=\frac{\pi^2}{p^2}\sum_{n=1}^\infty (-1)^{n+1} n^2\exp\left(-\frac{\pi^2 n^2 t}{2p^2}\right),\quad \text{as }\epsilon\to 0^+.
\end{align}
We obtain the distribution $\mu_A(p,x)$ of the endpoint by integration of the free propagator in $x$-direction weighted by $f_A(t)$ with respect to time $t$:
\begin{align}
 \mu_A(p,x) &= \int_{0}^\infty\frac{\diff t}{\sqrt{2\pi t}}\, \exp\left(-\frac{x^2}{2t}\right)f_A(t)\nonumber\\
 &= \frac{\pi}{p}\sum_{k=1}^\infty (-1)^{k+1} k\, e^{-\pi k|x|/p}
  = \frac{\pi}{4p}\frac{1}{(\cosh (\pi x/2p))^2}
\end{align}
Periodisation in $x$ with period $2\pi$ yields the endpoint distribution on the cylinder $\Lambda_2(p,x)$ \eqref{eqn:brownian}.

\subsection{Mixed boundary conditions}
\label{sec:mbc}
Consider the same problem as above, but impose reflecting boundary conditions at $y=0$ and absorbing boundary conditions at $y=p$. Let $P_M(y,t)$ the propagator in $y$-direction, defined via $P_M(y,t)\diff y = \mathsf{P}_{y_0=0}[y_t\in [y,y+\diff y]]$. It is solution of the diffusion equation $\partial P_M(y,t)/\partial t =1/2\, \partial^2 P_M(y,t)/\partial y^2$ and given by
\begin{equation}
  P_M(t,y)= \frac{2}{p}\sum_{n=0}^\infty \cos\left(\frac{(2n+1)\pi y}{2p}\right)\exp\left(-\frac{(2n+1)^2\pi^2t}{8p^2}\right).
\end{equation}
As previously, we compute the exit-time distribution from the probability current at $y=p$:
\begin{equation}
  f_M(t) = \frac{\pi}{2p^2}\sum_{n=0}^\infty (-1)^n (2n+1)\exp\left(-\frac{(2n+1)^2\pi^2t}{8p^2}\right).
\end{equation}
Consequently the distribution of the exit point at the upper boundary of the strip reads
\begin{align}
  \mu_M(p,x) &= \int_{0}^\infty\frac{\diff t}{\sqrt{2\pi t}}\, \exp\left(-\frac{x^2}{2t}\right)f_M(t)\nonumber\\
  {}&=\frac{1}{p}\sum_{n=0}^\infty (-1)^n\exp\left(-\frac{(2n{+}1)\pi |x|}{2p}\right)= \frac{1}{2p}\frac{1}{\cosh (\pi x/2p)}
\end{align}
Finally, periodisation with respect to $x$ with period $2\pi$ leads to $\Lambda_4(p,x)$ for the cylinder $\mathbb{T}_p$ \eqref{eqn:reflbrownian}.

\section{Endpoint distribution for hexagonal lattices}
\label{sec:lattice}
The purpose of this appendix is to give some idea about finite-size corrections to the scaling limit on the cylinder. Here we evaluate the discrete equivalent to \eqref{eqn:brownian} and compute the leading correction due to lattice effects. In principle, we dispose of a great variety for the choice of the lattice. Finite-size corrections not only turn out to be dependent on the choice of geometry, but on the lattice type as well. We shall content ourselves with the honeycomb lattice -- the most common choice.

\begin{figure}[htpb]
  \begin{tabular}{cc}
    \includegraphics[width=0.45\textwidth]{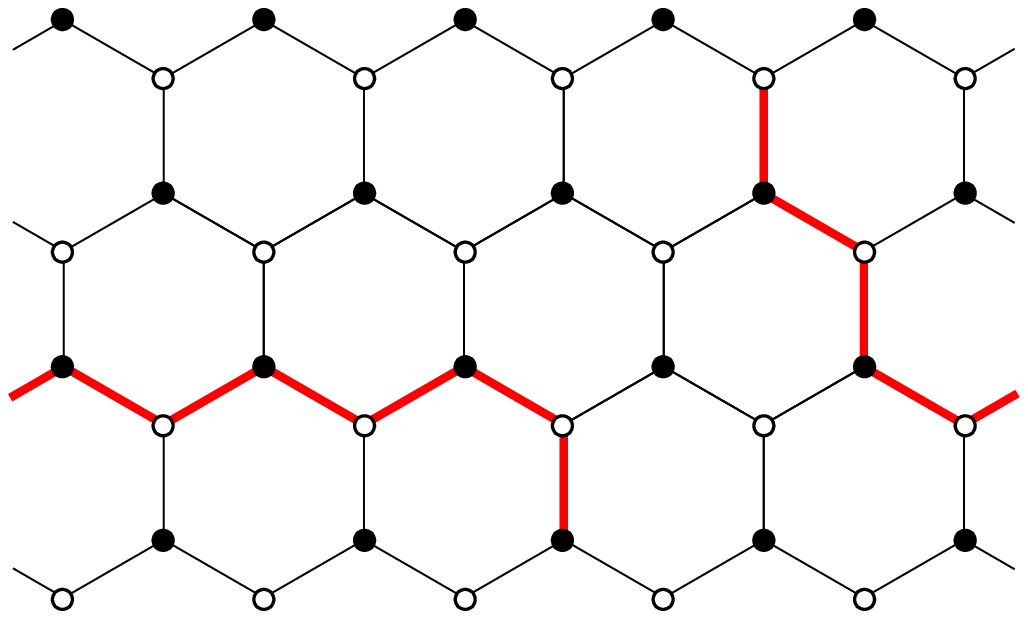}&
    \includegraphics[width=0.45
    \textwidth]{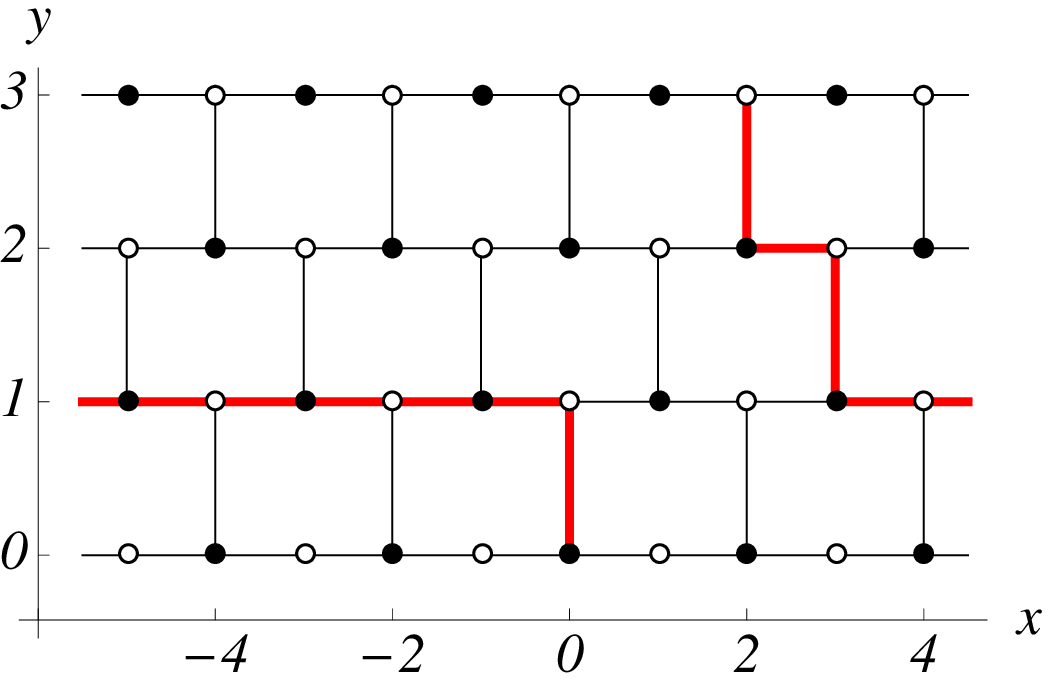}\\
    (a)&(b)
  \end{tabular}
  \caption{(a) Cylindrical lattice domain of $5\times 3$ honeycombs with walk from lower to upper boundary. The highlighted curve represent some walk from the lower to the upper boundary (b) Coordinate system for equivalent ``brick wall'' lattice.}
  \label{fig:geometry}
\end{figure}
We consider a finite lattice tube of $m\times l$ honeycombs, as shown in figure \ref{fig:geometry}a. We decompose the honeycomb lattice into two triangular sub-lattices whose sites are coloured by $\bullet$ and $\circ$. Nearest neighbours of a $\bullet-$site thus belong to the $\circ-$lattice and vice versa. We study lattice walks starting from a $\bullet-$site at the lower boundary to some any $\circ-$site at the upper boundary and shall impose absorbing boundary conditions on $\circ-$sites at either boundary. In fact, it shall be convenient to map the problem to a ``brick wall'' lattice as shown on figure \ref{fig:geometry}b, and introduce a suitable coordinate system.
Then it is straightforward to write the the master equations for the probability $P_{\bullet/\circ}(x,y;t)$ that the walker starting from $(0,0)$ can be found at $(x,y)$ at time $t$:
\begin{align}
 P_{\circ}(x,y;t+1)&=\frac{1}{3}\left(P_{\bullet}(x+1,y;t)+P_{\bullet}(x-1,y;t)+P_{\bullet}(x,y-1;t)\right)
  \label{eqn:master1}\\
P_{\bullet}(x,y;t+1)&=\frac{1}{3}\left(P_{\circ}(x+1,y;t)+P_{\circ}(x-1,y;t)+P_{\circ}(x,y+1;t)\right)
  \label{eqn:master2}
\end{align}
with boundary conditions $P_\circ(x,y=0;t)=P_\circ(x,y=l;t)=0$ and $P_\bullet(x,y;t=0)=\delta_{x,0}\delta_{y,0}$ where $\delta_{i,j}$ denotes the Kronecker symbol which takes the value $1$ if $i=j$, and $0$ otherwise. Moreover, identification of  $x$ and $x+2m$ leads to periodicity $P_\circ(x+2m,y;t)=P_\circ(x,y;t)$, $P_\bullet(x+2m,y;t)=P_\bullet(x,y;t)$. We shall need the local occupation times $G_{\bullet/\circ}(x,y)=\sum_{t=0}^\infty P_{\bullet/\circ}(x,y;t)$. The lattice walker is stopped if it walks from some $\bullet-$site at $y_e=1$ or $y_e=l-1$ to a $\circ-$site at $y_e=0$ or $y_e=l$ respectively, what happens with probability $1/3$. Therefore, the exit probability at the upper boundary is given by
\begin{equation}
  \mathsf{P}[x_e=x,y_e=l]= \frac{1}{3}G_{\bullet}(x,l-1), 
  \label{eqn:exit}
\end{equation}
where $x_e$ denotes the (random) horizontal position of the exiting walker. Notice that $x_e$ only takes odd/even integer values for even/odd $l$. Combining (\ref{eqn:master1}, \ref{eqn:master2}), it is possible to eliminate $G_{\circ}(x,y)$ and find the the second-order difference equation
\begin{align}
  G_{\bullet}(x,y)=&\frac{1}{6}\bigl(G_{\bullet}(x+2,y)+G_{\bullet}(x+1,y+1)+G_{\bullet}(x+1,y-1)\nonumber \\
  &G_{\bullet}(x-2,y)+G_{\bullet}(x-1,y+1)+G_{\bullet}(x-1,y-1)\bigr),
  \label{eqn:bulk}
\end{align}
for $0<y<l-1$. Using the boundary conditions, we furthermore find the equations
\begin{align}
  G_{\bullet}(x,0)&=\frac{1}{8}\bigl(G_{\bullet}(x+1,1)+G_{\bullet}(x-1,1)\bigl)+\frac{9}{8}\,\delta_{x,0}
  \label{eqn:bdzero}
  \\
  G_{\bullet}(x,l-1)&=\frac{1}{7}\bigl(G_{\bullet}(x+1,l-2)+G_{\bullet}(x+2,l-1)
  \nonumber\\
  &\qquad+G_{\bullet}(x-1,l-2)+G_{\bullet}(x-2,l-1)\bigr)
  \label{eqn:bdl}
\end{align}
Solving \eqref{eqn:bulk}, \eqref{eqn:bdzero} and \eqref{eqn:bdl} all together leads to a rather lengthy and tedious calculation. The reader may find a detailed account of the strategy with applications to other lattice types in \cite{henri:03}. Here we only sketch the solution.

First, we solve the bulk equation \eqref{eqn:bulk} by a separation ansatz $G_{\bullet}(x,y)= P(x)Q(y)$. Forming linear combinations of solutions found from this ansatz, we obtain the general bulk solution
\begin{align}
  G_{\bullet}(x,y)=&A_0+C_0 y+(A_{m}+C_{m}y)(-1)^{x+y}\label{eqn:bulkgeneral}\\
  &+\sum_{k\in I_1}(A_k e^{i\alpha_k x}+B_k e^{-i\alpha_k x})(e^{\gamma_k y}+C_k e^{-\gamma_k y})\nonumber \\
  &+\sum_{k\in I_2}(-1)^y(A_k e^{i\alpha_k x}+B_k e^{-i\alpha_k x})(e^{\gamma_k y}+C_k e^{-\gamma_k y})
  \nonumber
\end{align}
where $A_i$, $B_i$ and $C_i$ are constants, $I_1=\{k\in \mathbb{Z}\mathop{|}0<k<m/2 \vee 3m/2<k<2m\}$ and $I_2=\{k\in \mathbb{Z}\mathop{|} m/2<k<3m/2,\,k\neq m/2\}$ index sets, and
\begin{equation}
  \alpha_k = \frac{\pi k}{m}, \quad \cosh \gamma_k = \frac{3-\cos 2\alpha_k}{2|\cos\alpha_k|}\quad \text{and}\quad k=0,1,\dots,2m-1
\end{equation}
However, we exclude $k= m$. Moreover, for some technical reason the formula only is valid as long as $m$ is odd. The remaining constants are determined by matching \eqref{eqn:bulkgeneral} to \eqref{eqn:bdzero} and \eqref{eqn:bdl} what yields
\begin{align}
  A_k=B_k &= \frac{1}{2m |\cos\alpha_k|\sinh(\gamma_k l)},\,
  C_k= -e^{2\gamma_k(l-1)}\frac{1+2e^{\gamma_k}|\cos\alpha_k|}{1+2e^{-\gamma_k}|\cos\alpha_k|},\nonumber
\end{align}
and $A_0=A_{m}=(3l-1)/(4ml)$, $C_0=C_{m}=-3/(4ml)$. Putting all pieces together, we find the solution for \eqref{eqn:exit}. However, if we condition the walk to exit at the upper boundary, we still have to divide by the probability $\mathsf{P}[y_e=l]$ what amounts to a normalisation. After some algebra we obtain the final result (for odd $m$)
\begin{align}
  \mathsf{P}[x_e=x|y_e=l]=&\frac{1}{m}+\sum_{k=1}^{(m-1)/2}\frac{2l\,\cos (\alpha_kx) \sinh\gamma_k}{m \sinh\gamma_kl}.
\end{align}
Having this lattice result, it is interesting to compute its scaling limit and study corrections. In the case where $m$ is odd, $x$ takes even values from $-(m-1)$ to $m-1$. Introducing a lattice scale $a$ such that $x=\xi/a$, $\xi\in \mathbb{R}$, we would like to study the scaling limit $a\to 0^+$ , $m,l\to +\infty$ in such a way that the geometrical height $p=\sqrt{3}(l-1/3)a$ and width $2\pi=2am$ remain finite. Therefore $l/m=p/\pi - a/(\sqrt{3}\pi)$.
\begin{align}
   \lambda(p,\xi,a)& = \frac{1}{2a}\mathsf{P}[x_e=x|y_e=l]
   =\frac{\sqrt{3}\,l}{\pi m}\sum_{k=-\infty}^\infty \frac{ k\,\cos k\xi}{\sinh (\sqrt{3}\,(l/m)k)}\nonumber\\
  & =\frac{1}{4\sqrt{3}(l/m)}\sum_{n=-\infty}^\infty\frac{1}{\left[\cosh\left(\frac{(x+2\pi n)}{2\sqrt{3}(l/m)}\right)\right]^2}\nonumber \\
  &=\left(1+\frac{a}{\sqrt{3}p}\right)\lambda_2(p,\xi)+O(a^2)
\end{align}
We see that the first correction to the scaling limit is completely naturally proportional to the lattice scale $a$.

\section{Analysis in the bulk} 
\label{sec:bulk}

This appendix collects some facts about bulk properties for SLE on the cylinder. Let us denote the probability, that the trace $\gamma$ passes to the left of some given point $z$ on the covering space by $\omega(p,z,\overline{z})$. It is solution of the diffusion equation
\begin{equation}
\frac{\partial \omega(p,z,\overline{z})}{\partial p}= H(p,z)\frac{\partial \omega(p,z,\overline{z})}{\partial z}+H(p,\overline{z})\frac{\partial \omega(p,z,\overline{z})}{\partial \overline{z}}+\frac{\kappa}{2}\left(\frac{\partial}{\partial z}+\frac{\partial}{\partial \overline{z}}\right)^2\omega(p,z,\overline{z})
  \label{eqn:fpomega}
\end{equation}
with boundary conditions $\omega(p,z,\overline{z})\to 0$ as $\text{Re}\,z\to -\infty$ and $\omega(p,z,\overline{z}) \to 1$ as $\text{Re}\,z\to +\infty$. The derivation of \eqref{eqn:fpomega} is similar to \eqref{eqn:averagebm} and \eqref{eqn:pde}, using an infinitesimal argument. Using the relationship $H(p,z) = v(p,z-ip)-i$ to the velocity field defined above and \eqref{eqn:pde4v}, it is not difficult to show that $H(p,z)$ is a Burgers flow, too. We have
\begin{equation}
 \frac{\partial H(p,z)}{\partial p} = H(p,z)\frac{\partial H(p,z)}{\partial
  z}+\frac{\partial^2H(p,z)}{\partial z^2}
  \label{eqn:burgers4h}
\end{equation}
Using the Cole-Hopf transformation \eqref{eqn:colehopf}, we may write  $H(p,z)=\partial U(p,z)/\partial z$ with $U(p,z)=2\log \epsilon(p,z-ip)-iz -  p/2$.

For $\kappa=2$ as $\text{Im}\, z\to p^-$ we require that
$\omega(p,z,\overline{z})$ tends to $\lambda_2(p,x)$ (with $x=\text{Re}\,z$), and $\omega(p,z,\overline{z})\to \Theta(x)$ as $\text{Im}\, z\to 0^+$. Although we have not found the solution with these boundary conditions, we have determined a special solution $\tilde{\omega}(p,z,\overline{z})$ with a periodised version of these boundary conditions which has interesting relations to dipolar SLE$_2$. We ask that for $\text{Im}\, z\to p^-$ it becomes $\tilde{\omega}(p,z,\overline{z}) = \int_0^x\diff x\, \Lambda_2(p,x) = (p\, v(p,x)+x)/2 \pi$, and that for $\text{Im}\, z\to 0^+$ it reproduces the quasi-periodic step function $\tilde{\omega}(p,z,\overline{z})=\lim_{n\to\infty}\sum_{k=-n}^n(\text{sign}\, (x-2\pi k))/2$.

Obviously, both $H(p,z)$ and $H(p,\overline{z})$ are solutions to \eqref{eqn:fpomega}, however with a simple pole at $z=0$.
Inspired from the boundary condition at $\text{Im}\,z=p$, we shall consider the function $f(p,z,\overline{z}) = (\text{Im}\,z)(H(p,z)+H(p,\overline{z}))$, which has no pole. However, using the differential operator $\mathcal{A}$ defined as
\begin{equation}
\mathcal{A}= \left[\frac{\partial}{\partial p}- H(p,z)\frac{\partial}{\partial z}-H(p,\overline{z})\frac{\partial }{\partial \overline{z}}-\left(\frac{\partial}{\partial z}+\frac{\partial}{\partial \overline{z}}\right)^2\right]
\end{equation}
we see that $\mathcal{A}f(p,z,\overline{z})= - (H(z)^2-H(\overline{z})^2)/2i$. Hence we should find some solution to $\mathcal{A}g(p,z)=iH(z)^2/2$ without poles (since this would again lead to a singular solution). Hence we have to search for a suitable $g(p,z)$ with at most a logarithmic singularity at $z=0$. This can be done by observing that $U(p,z)=2\log \epsilon(p,z-ip)-iz - p/2$ is solution to the equation
\begin{align}
  \frac{\partial U(p,z)}{\partial p} &= \frac{1}{2}\left(\frac{\partial U(p,z)}{\partial z}\right)^2 + \frac{\partial^2 U(p,z)}{\partial z^2}\nonumber\\
  &= H(p,z)\frac{\partial U(p,z)}{\partial z}+\frac{\partial^2 U(p,z)}{\partial z^2}-\frac{1}{2}\,H(p,z)^2.
\end{align}
as can be seen from Burgers equation for $H(p,z)$ \eqref{eqn:burgers4h}. Hence we have found a solution
$
\tilde{\omega}(p,z,\overline{z}) = C_1+C_2(f(p,z,\overline{z})- i (U(p,z) - U(p,\overline{z}))),
$
where $C_1,\,C_2$ are constants. It is a solution of \eqref{eqn:fpomega} without poles for $0\leq \text{Im}\,z\leq p$. The choice $C_1=1/2$ and $C_2=1/4\pi$ leads to the desired boundary conditions for $\tilde{\omega}(p,z,\overline{z})$:
\begin{equation}
  \tilde{\omega}(p,z,\overline{z}) = \frac{1}{2}+\frac{\text{Im}\,z}{4\pi}(H(p,z)+H(p,\overline{z}))+\frac{i}{4\pi}(U(p,z)-U(p,\overline{z})) \label{solucyl} 
\end{equation}
$\tilde{\omega}(p,z,\overline{z})$ is related to $\omega(p,z,\overline{z})$ by periodisation
\begin{equation}
  \tilde{\omega}(p,z,\overline{z}) = \lim_{n \to \infty} \sum_{k=-n}^{n} \left(\omega(p,z + 2 \pi k,\bar z + 2 \pi k ) - \frac{1}{2} \right) 
\end{equation}
i.e. we have only found a (quasi-)periodic version of $\omega(p,z,\bar z)$. 

The dipolar limit $p\to 0^+$ turns out to be quite interesting. In that limit the
quasi-periodicity is irrelevant and one obtains the result for dipolar SLE$_2$.
One can replace in that limit $H(p,z) \to H_{ss}(p,z)=\frac{\pi}{p} \coth \pi z/2p
- z/p$, the single shock expression, and the corresponding expression $2 \log \epsilon(p,z)\sim -z^2/(2p) + 2 \log \cosh \pi z/2p$, up to a $p$-dependent unimportant constant. One then easily obtains from (\ref{solucyl}):
\begin{equation}
\tilde{\omega}(p,z,\overline{z})+\frac{1}{2}\sim\omega_d(p,z,\overline{z})= 1-\frac{1}{\pi}\,\text{Im}\log\sinh\frac{\pi z}{2p}+\frac{\text{Im}\,z}{4 p}\left(\coth \frac{\pi z}{2p}+\coth\frac{\pi \overline{z}}{2p}\right)   \label{omegadip} 
\end{equation}
This is nothing but the probability that dipolar SLE$_2$ passes to the left of a given point $z$ within a strip of height $p$. It satisfies the general equation:
\begin{equation}
\left[ \frac{\pi}{p} \coth \frac{\pi z}{2p} \frac{\partial}{\partial z}  + \frac{\pi}{p} \coth \frac{\pi \overline{z}}{2p} \frac{\partial}{\partial \overline{z}} + \frac{\kappa}{2} \left(\frac{\partial}{\partial z}+\frac{\partial}{\partial \overline{z}}\right)^2 \right] \omega_d(p,z,\overline{z}) = 0 \label{omegadeq} 
\end{equation}
studied in \cite{bauer:05} where solutions could be found only for $\kappa \geq 4$ (in the
form of simple harmonic functions). Here we recover, from a limit of a more general object defined on $\mathbb{T}_p$, a result for $\kappa=2$ obtained only very recently by a rather different method \cite{bauer2:08}. The
fact that (\ref{omegadip}) satisfies both (\ref{omegadeq}) (an equation with no $\partial_p$ terms) and
the dipolar (i.e. single shock) limit of (\ref{eqn:fpomega}) (an evolution equation as a function of $p$) is easily understood by noting that it also satisfies $(p \partial_p + z \partial_z + \bar z \partial_{\bar z} )\omega_d(p,z,\bar z)=0$ which expresses the dilatation invariance of the dipolar SLE, explicitly broken by the period of the cylinder.

One may wonder whether there is a way to extend the known harmonic solutions \cite{bauer:05} of the dipolar limit, to the full cylinder. Here we only present the case $\kappa = 4$ which can be constructed by integration of the result given in section \ref{sec:kappa4}. In fact, this amounts to periodise the result from dipolar SLE for a strip of height $p$
\begin{equation}
\omega_d(p,z,\overline{z}) =1- \frac{1}{\pi}\,\text{Im}\log \tanh \frac{\pi z}{4p}
\end{equation}
in an appropriate way. Symmetric periodisation and leads to
\begin{equation}
  \tilde\omega(p,z,\overline{z}) = \,\lim_{n\to \infty}\sum_{k=-n}^n\left(\frac{1}{2}-\frac{1}{\pi}\,\text{Im} \log \tanh\left( \frac{\pi (z+2\pi k)}{4p}\right)\right)
\end{equation}
which is a harmonic solution to \eqref{eqn:fpomega} with required boundary conditions at $z=x+ip$
\begin{equation}
  \tilde\omega(p,x+ip,x-ip)=\int_0^x\diff x\, \Lambda_4(p,x).
\end{equation}

\section{Path integral for general $\kappa$}
\label{apppath}

Let us indicate how one can take advantage of \eqref{eqn:psi} to express the solution of (\ref{eqn:pde})
as a path integral for general $\kappa$. We may write
\begin{equation}
  \psi(p,x)=\int_{-\infty}^\infty\mbox{d}y\,K(p,x;y)\psi(0,y)
\end{equation}
where $K(p,x;y)$ is the Euclidean propagator for \eqref{eqn:psi}. It given by a path-integral representation
\begin{equation}
  K(p,x;y) = \int_{x(0)=y}^{x(p)=x}[\mbox{d}x(t)]\,e^{-S[x(t)]}
\end{equation}
where the action is defined as
\begin{equation}
S[x(t)]=\int_0^p\mbox{d}t\left(\frac{\dot{x}(t)^2}{2\kappa}
+\frac{\kappa-2}{2\kappa}\frac{\partial v(t,x(t))}{\partial x}\right).
\end{equation}
Hence, we find a nice path-integral representation \cite{ginanneschi:97,ginanneschi:98} for the general solution $\omega(p,x)$ of (\ref{eqn:proba}):
\begin{equation}
  \omega(p,x) = \epsilon(p,x)^{-2/\kappa}\int_{-\infty}^\infty\mbox{d}y\,
  \epsilon(0,y)^{2/\kappa} \omega(0,y)\int_{x(0)=y}^{x(p)=x}[\mbox{d}x(t)]\,e^{-S[x(t)]}
\end{equation}
which reproduces the exact solution (\ref{eqn:omega}) in the case $\kappa=2$. It may be
useful for perturbation theory around $\kappa=2$, using the Fourier series decomposition:
\begin{equation}
  v(p,x) = \sum_{n=1}^\infty \frac{4 e^{- n p} }{1-e^{-2 n p}} \sin(n x) 
\end{equation}
if one treats the apparent singular behaviour of $\epsilon(0,y)^{2/\kappa}$ for
$\kappa \neq 2$ (e.g. shifting the integration from $0$ to small $p_0$ and using the known dipolar result for $\omega(p_0,x)$).

\end{document}